\documentclass[12pt]{article}
\usepackage[utf8]{inputenc}
\usepackage[a4paper]{geometry}
\usepackage{graphicx}
\usepackage{fourier-orns}
\usepackage{amsmath}
\usepackage{authblk}
\usepackage{hyperref}
\usepackage{float}
\usepackage[table]{xcolor}
\usepackage{longtable}
\usepackage{multirow}
\usepackage{color}
\usepackage[compact]{titlesec}
\usepackage{booktabs}
\usepackage{lipsum}

\usepackage{graphicx}
\graphicspath{ {./images/} }

\titlespacing*{\subsubsection}
  {0pt}{0.5\baselineskip}{0\baselineskip}

\title{
The UK Universities Superannuation Scheme valuations 2014-2023: gilt yield dependence, self-sufficiency and metrics}
\author{Jackie Grant \thanks{j.j.grant@sussex.ac.uk} }
\date{\today}

\affil{\textit{Department of Physics and Astronomy, University of Sussex, Brighton BN1 9QH, U.K.}}

\begin{document}

\maketitle

\begin{abstract}
This review considers the Universities Superannuation Scheme (USS) valuations from 2014 to 2023. USS is a \pounds 70-80 billion Defined Benefit pension scheme with over 500,000 members who are employed (or have been employed) at around
70 UK universities. Disputes over USS have led to a decade of industrial action.
New results are presented showing the high dependence of USS pension contributions on the return from UK government bonds (the gilt yield). The two conditions of the USS-specific `self-sufficiency' (SfS) definition are examined. USS data are presented along with new analysis. It is shown that the second SfS condition of `maintaining a high funding ratio' dominates USS modelling to amplify gilt yield dependence, inflating the SfS liabilities beyond the regulatory requirements, and leading to excessive prudence. 
The Red, Amber and Green status of USS metrics `Actual' and `Target' Reliance are also examined. It is shown that Target Reliance tethers the cost of future pensions to the SfS definition and that Actual Reliance  can simultaneously be Green and Red. 
Implications for regulatory intervention are considered. 
An aim of this review is to support evidence-based decision making and 
consensus building. 

\end{abstract}

\newpage

\section{Introduction}

The Universities Superannuation Scheme (USS) is a {\pounds}70-80 billion  UK pension scheme for hundreds of thousands of staff at around 70 universities and 270 smaller academic and related institutions. Over the last decade USS has carried out five  full valuations to satisfy regulatory requirements. Of these, the 2017 valuation led employers to propose closing the scheme. Following unprecedented strike action the closure was cancelled and the scheme remains open \cite{UCU2017USS,FATest12017}. The 2020 valuation saw employers impose significant cuts and increased costs \cite{grant2022distribution}. But (against the backdrop of even stronger industrial action) agreement was reached within 18 months to reverse and compensate the cuts and significantly reduce costs \cite{UUK_UCU_JS_Oct_2023}. 

A central focus of disputes over USS has been the high costs proposed for pension benefits and the volatility of those costs. This paper surveys the last decade of valuation data to investigate the causes. The  results are startling. 

Section \ref{sec:gilt_yield}, pp3-11, demonstrates the very high \textbf{gilt yield dependence} of costs. USS data show that 95-99\% of the volatility in costs can be attributed to movements in the return on UK Government bonds\footnote{See Appendices for Glossary \ref{glossary}.}. This is difficult to explain for a scheme that is invested in a diversified global portfolio of 60\% equities. 

The \textbf{USS self-sufficiency definition} is examined in Section \ref{sec:self sufficiency}, pp12-22. USS data and preliminary independent analysis are presented. Both suggest that a little-known USS-specific `funding ratio' condition used in USS's modelling is inflating self-sufficiency liabilities. Secondly, this funding ratio does not predict the ability to pay benefits. It is not clear what the funding ratio condition is  usefully doing. An absence of USS analysis on the funding ratio is noted. 

The \textbf{USS Target and Actual Reliance metrics}, which can be Red, Amber or Green are explored in Section \ref{sec:ActTarRel}, pp23-26. It is shown that Actual Reliance can be simultaneously Green and Red. It is also demonstrated that Target Reliance tethers the Technical Provisions to the self-sufficiency liabilities. This presents an explanation for the 97-99\% correlation between gilt yield and the USS pre-retirement discount rate (for 90\% equities) that is absent in their expected returns. 

Section \ref{sec:IRMF}, pp27–29, considers the \textbf{USS Integrated Risk Management Framework} (IRMF) that underpins the  valuation. It is suggested that the self-sufficiency funding ratio condition is driving the entire IRMF, and thereby inflating self-sufficiency liabilities, which in turn invites regulatory intervention. 

A \textbf{Summary \ref{sec:summary} and Conclusions \ref{sec:concl}}, pp30-32, provide a synopsis with the aim of supporting evidence-based future valuations and informed decision making. 

\newpage

\section{The gilt yield determines USS contributions}\label{sec:gilt_yield}

This section shows that over the period 2014-2023
the amount charged by USS to employers and employees 
for pension benefits is highly dependent on the gilt yield (the annual percentage yield 
on UK Government bonds). 

\begin{figure}[hb!]

\centering

\includegraphics[width=0.9\textwidth]{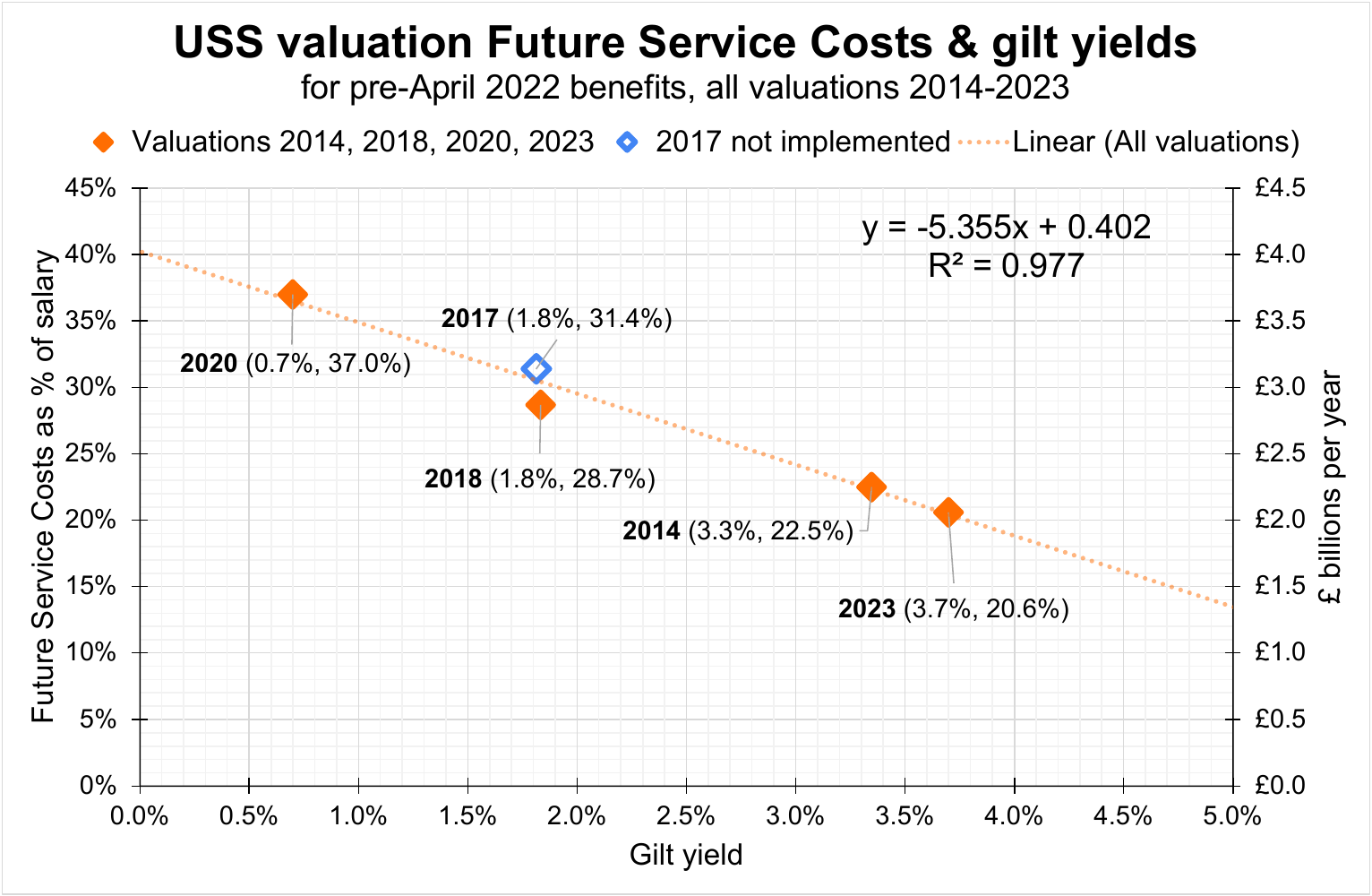}

  \caption{The five USS valuations 2014-2023 show a 98\% correlation, $R^2$, between Future Service Costs and gilt yield. The FSCs are shown as percentage of salary (left-axis) and as \pounds billion in annual cost to the sector (right-axis). All FSCs are for pre-2022 benefits (the benefit structure before cuts were imposed in April 2022). 
  }
    \label{fig:gilt corr val 14-23}

\end{figure}

\begin{figure}[hb!]

\centering

\includegraphics[width=0.9\textwidth]{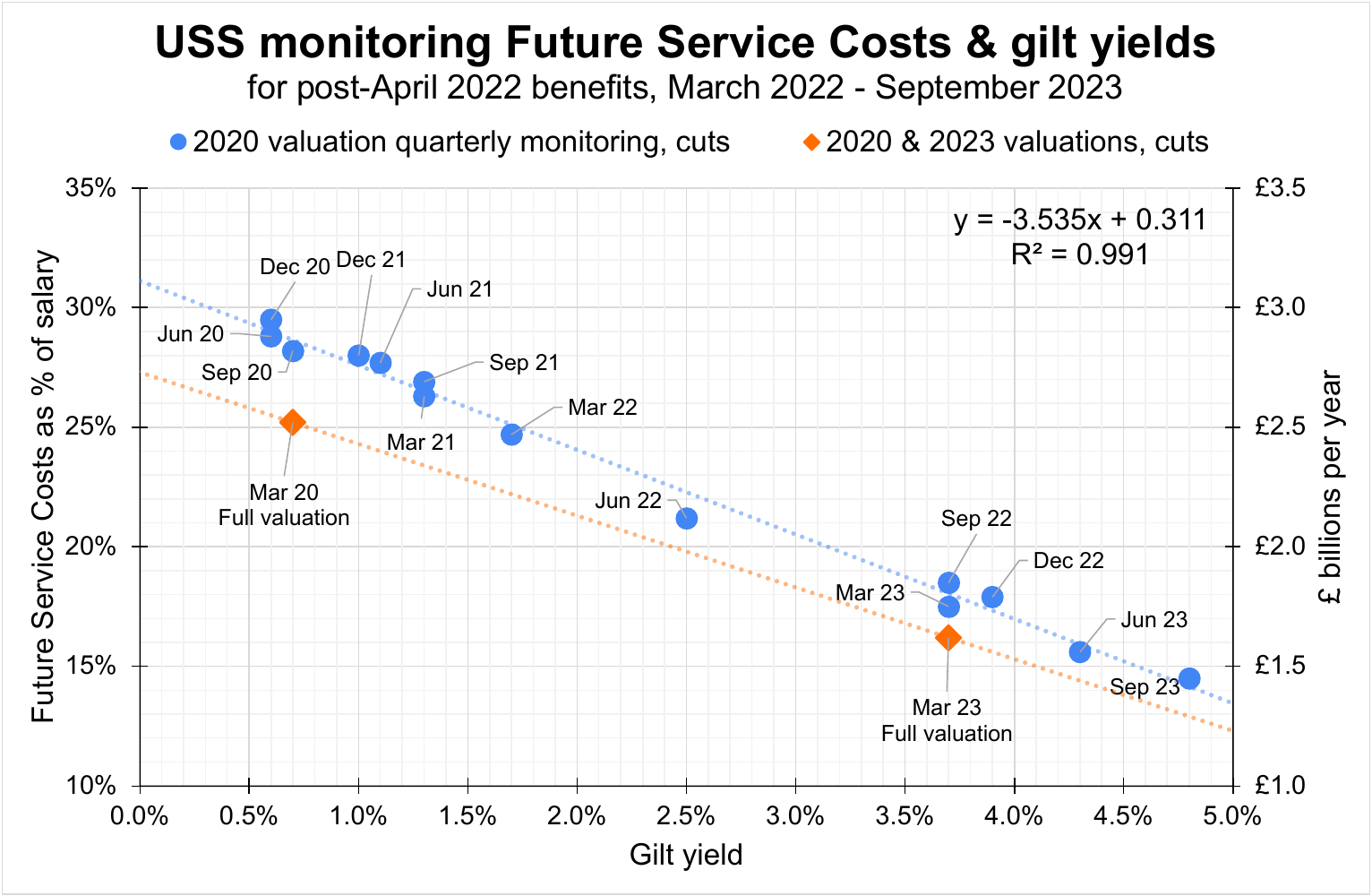}

  \caption{USS quarterly monitoring of the 2020 valuation to September 2023 for post-2022 benefits (with cuts). A correlation, $R^2$, of 99.1\% is found between FSCs and the gilt yield. The monitoring is slightly more prudent but demonstrates the same features seen in valuations 2014-2023 as detailed in Figures \ref{fig:gilt corr val 14-23} and \ref{fig:gilt corr mon 2020}. 
  }

    \label{fig:gilt corr mon  2020 cuts}

\end{figure}

\subsection{Future Service Cost gilt yield dependence}

Future Service Costs (FSCs) are the USS-calculated costs per year of providing \textit{future}\footnote{USS also requires payment for their estimation of any past deficits, see section \ref{subsec:TP DR gilt yield}.} pensions to over 200,000 university staff enrolled in the scheme. The costs are usually expressed as a percentage of employee's salary, or equivalently employers' annual payroll, and referred to as `contribution rates'. 
The costs are then shared between employers and employees in the ratio of around 70\% to 30\%\footnote{See `contributions' and `cost-sharing' in Glossary \ref{glossary}.}. 
Figure
\ref{fig:gilt corr val 14-23}
shows the five USS valuations 2014--2023, and demonstrates a very high correlation between FSCs and the gilt yield at the time of the valuation. Figure \ref{fig:BoE gy corr 2000-2024} in Appendix \ref{app:gilt yield} presents the gilt yield 2000-2024.

\begin{table}[hb!]
    \centering
\begin{tabular}{|r|r|r|r|}

\hline
\textbf{Type of USS data }& \textbf{Benefits}& \textbf{Data source}   & \textbf{$R^2$}  \\
\hline
\hline
 
 All valuations 2014 - 2023  & pre-2022  & USS \& BoE & 97.6\%  \\
\hline

Monitoring Mar 2020 - Sep 2023 & post-2022 &  USS & 99.1\% \\
\hline

Monitoring Mar 2020 - Sep 2023 & pre-2022  &  USS & 95.5\%\\
\hline
\hline

Monitoring Mar 2020 - Oct 2022  & post-2022  & Fig. \ref{fig:BillGraph2022} \& BoE & 87.8\% \\
\hline

Monitoring Mar 2018 - Feb 2020 & pre-2022  &  Fig. \ref{fig:BillGraph2022} \& BoE & 85.6\% \\
\hline

Monitoring Mar 2014 - Feb 2017  & pre-2022   & Fig. \ref{fig:BillGraph2022} \& BoE & 89.6\%\\
\hline

\end{tabular}

\caption{The first three rows show $R^2$ values for valuations, and quarterly monitoring since 2020 for post-2022 and pre-2022 benefits. The last three rows show monthly monitoring of the three implemented valuations of 2014, 2018 and 2020, using a data grab from Figure \ref{fig:BillGraph2022} and Bank of England (BoE) 20-year gilt yields.}
    \label{tab:all_R2_val_mon}
\end{table}

In 2020 the gilt yield was 0.7\%, and FSCs 37\%. In 2023 the gilt yield was higher at 3.7\% and FSCs had reduced from 37\% to 20.6\%. The 2023 annual university sector payroll for those employees in USS was around \pounds 10 billion, so 37\% represents an annual cost of \pounds 3.7bn to employers and employees. 

To reduce these costs employers proposed significant pension cuts which they imposed in April 2022 \cite{grant2022distribution}. The result was widespread industrial action across UK universities. 
The 2023 valuation FSCs were much lower at 20.6\%. This 16.4 percentage point drop in FSCs due to a gilt yield increase of 3 percentage points meant a cost reduction of around \pounds 1.6 billion per year\footnote{This annual FSC reduction of \pounds1.6bn omits the deficit savings of \pounds0.6bn. See Section \ref{subsec:TP DR gilt yield}.}. In 2023, as pressure from  industrial action continued to grow, agreement was reached to fully reverse and compensate for the cuts imposed in 2022, and to implement this at April 2024 \cite{UUK_UCU_JS_Oct_2023}.  

To inform decision-making between valuations, USS uses monthly or quarterly post-valuation monitoring to indicate `direction of travel', but notes that the monitoring is different to the more detailed analysis used for full valuations \cite{USS_vals_mon}. The valuations 2014-2023 (Fig. \ref{fig:gilt corr val 14-23}) and 2020 quarterly monitoring (Fig. \ref{fig:gilt corr mon  2020 cuts} and \ref{fig:gilt corr mon  2020}) show some differences, in that the monitoring appears slightly more prudent, but the valuations and monitoring display a similarly high dependence on gilt yield. 

Table \ref{tab:all_R2_val_mon} considers all implemented valuations and post-valuation monitoring in the last decade. 
The first three rows for 2014-2023 valuations and 2020-2023 monitoring use USS public data and show a correlation of 95-99\%. 
The remaining rows use data grab software \cite{Rohatgi2022} applied to a USS graph of FSC monitoring for 2014-2023. The lower quality monitoring data from 2014 onward are compared to USS published data and the lower correlations (85-90\%)
are shown to be explained by the lower quality nature of the available data. See Appendix \ref{app:gilt yield} for discussion.

\subsection{FSC calculations from prudent return on equities }\label{sec:Pre-ret}

\begin{figure}[hb!]

\centering

\includegraphics[width=0.9\textwidth]{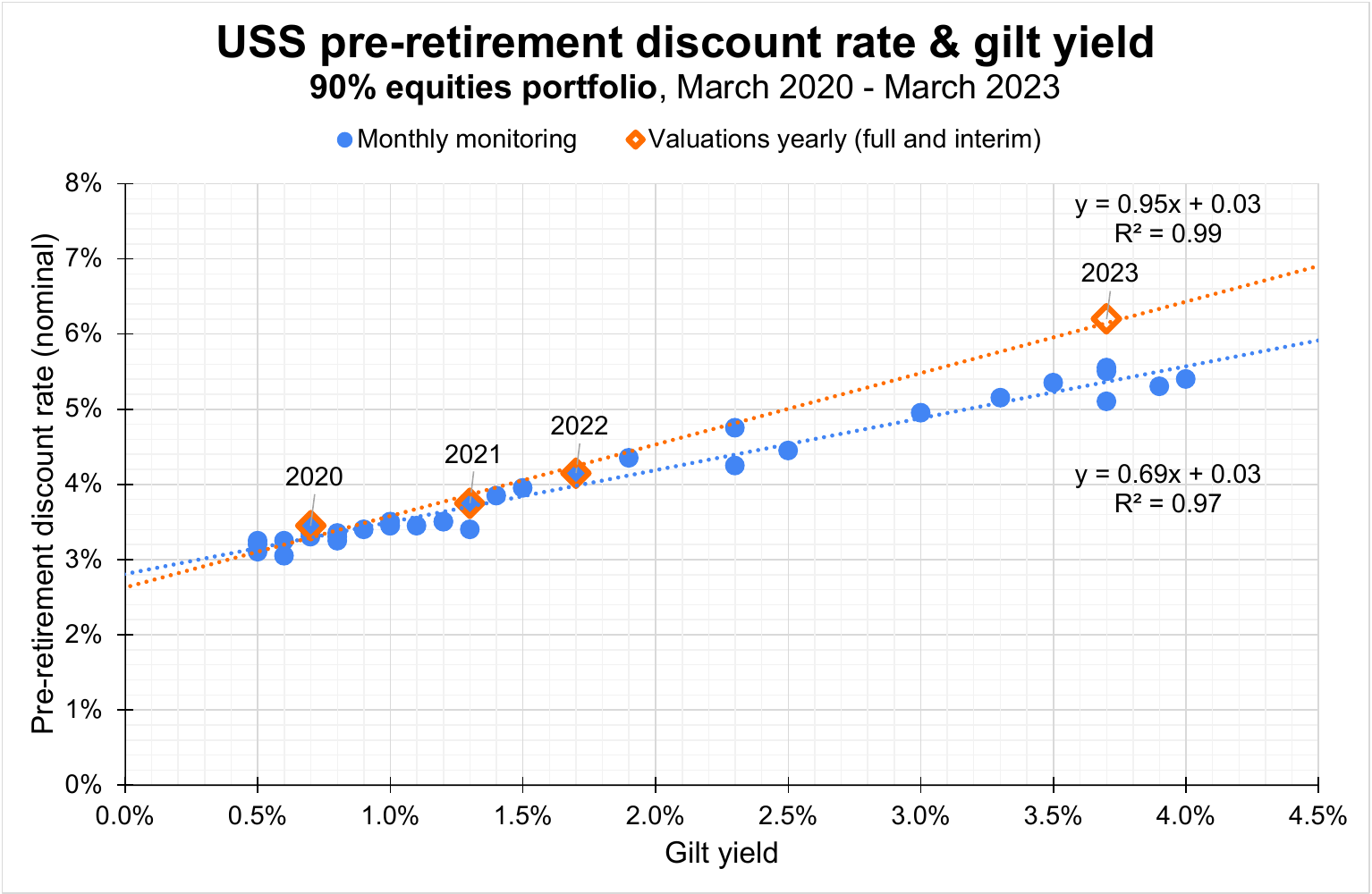}

  \caption{USS pre-retirement discount rate which is the prudently adjusted annual expected return on a 90\% equities portfolio. Monthly monitoring data since March 2020 are shown for annual full or interim valuations. The prudent returns show a 99\% and a 97\% correlation with the gilt yield. 
  }

    \label{fig:gilt corr pre-ret pru 2020-2023}

\end{figure}

Like all UK DB pension schemes, USS calculates the benefits that scheme members have accrued in the past and are projected to accrue in the future, and when these might need to be paid. These are based on a range of assumptions including  price increases, salary growth and life expectancy. USS publishes these as  `cashflows'. 

USS then applies a `discount rate' (DR) to these projected cashflows to estimate a present value of the projected benefits. There can be several choices of discount rate, for example it can be a best estimate discount rate, representing expected returns on the assets held, or `prudently' adjusted with a chosen percentage confidence level below their assumption for the best estimate of return. 

UK legislation requires that pension schemes apply some level of prudence when calculating the present value of projected promised benefits \cite{UKParl_PensionsAct_2005}. The present value of promised pensions, discounted at a prudent DR, is usually referred to as the Technical Provisions (TP). A similar prudence is usually applied to the calculation of FSCs for benefits expected to accrue in the future (although legislation does not require this) and USS calculates FSCs from a  prudent FSC DR \cite{10.1093/oso/9780198885962.001.0001}.

Since 2020 USS has adopted a dual discount rate (DDR) approach whereby a different rate is applied to benefits in payment (owed to those who have retired) and benefits yet to be paid (owed to those who have not yet retired). The USS DDR approach considers two hypothetical investment portfolios: the pre-retirement and post-retirement (pre- and post-ret) portfolios. 

USS states that their pre- and post-ret DRs are prudently adjusted from the best estimate of expected annual returns on these two portfolios.
The pre-ret portfolio is approximately 90\% equities and 10\% bonds, so represents high-growth long-term investments. The post-ret portfolio is approximately 10\% equities and 90\% bonds.  USS then sets the weightings for FSC DR as 55\% pre-ret DR plus 45\% post-ret DR\footnote{Weightings calculated from pp19-21 of USS 2023 TP consultation \cite{USS_Val_TP_SI_2023} details on github \cite{SussexUCUgithub}.} to represent the membership profile in terms of years to retirement. It is worth noting that these weighting produce an overall allocation of effectively 54\% equities for the FSCs investment portfolio, which appears to be lower than the 60\% allocation for the overall USS Valuation Investment Strategy\footnote{The Valuation Investment Strategy is discussed in the Glossary \ref{glossary}.}.

Figure \ref{fig:gilt corr pre-ret pru 2020-2023} shows the USS assumptions for pre-ret DR (or prudent annual nominal return on the pre-ret portfolio of 90\% equities) since March 2020. 
The monthly monitoring and yearly valuations show similar behaviour including a very high correlation (97\% to 99\%) with the gilt yield. The pre-retirement portfolio is 90\% equities so it is not clear how such a high dependence of the pre-ret DR on the UK government bond rate can be explained. 
Exploration of this dependence is provided by turning to analysis of the USS best estimate of expected returns.

\begin{figure}[hb!]

\centering

\includegraphics[width=0.9\textwidth]{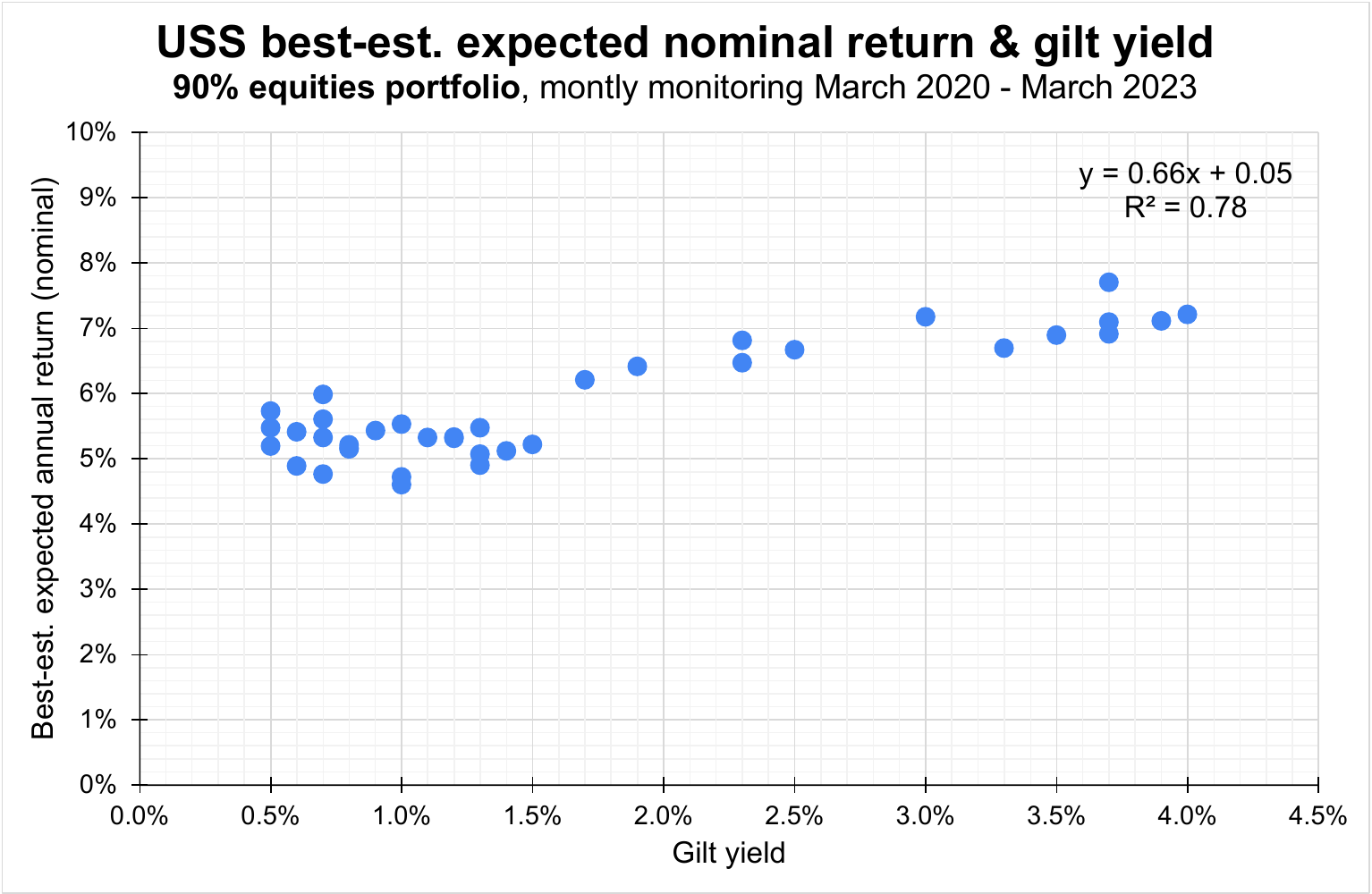}

  \caption{USS monitoring of best estimate returns on their pre-retirement investment portfolio of 90 \% equities and 10 \% bonds. Although there is some correlation with the gilt yield ($R^2=78\%$) it is notably much lower than the high correlations ($R^2=97-99\%$) seen in the prudently adjusted return (pre-retirement discount rate) of Figure \ref{fig:gilt corr pre-ret pru 2020-2023}. 
  Valuation data is omitted as it is not available. }
    \label{fig:pre-ret_unbiased}

\end{figure}

Figure \ref{fig:pre-ret_unbiased} shows USS monthly monitoring of their best estimate (or unbiased) return on the pre-retirement portfolio of 90\% equities, for March 2020-March 2023. The best estimates of returns show a much lower correlation, $R^2$=78\%, than the prudently adjusted returns (the pre-retirement discount rates) with correlations 97\% to 99\%.  It is remarkable that the prudently adjusted pre-ret DR show such a very high gilt yield correlation when the best estimate returns do not. It would seem that a mechanism is bypassing considerations of the best estimates of returns on equities and instead setting discount rates by a direct tethering to the gilt yield. Suggested mechanisms are considered in the Summary \ref{sec:summary}.

Next the volatility in costs, due to this gilt yield dependence of the pre-retirement discount rate, is considered.

At each valuation USS presents analysis showing sensitivity of FSCs to a range of assumptions. Table \ref{tab:pre ret sensitivity} shows the FSC sensitivity to changes in the pre-ret DR as presented by USS in both 2020 and 2023.

\begin{table}[h]
    \centering
\begin{tabular}{|l|r|r|}

\hline
 \textbf{}& \textbf{2020} &  \textbf{2023}   \\
\hline
\hline
USS reported FSC for pre-ret DR of gilts+2.5\% & 34.5\% &  20.6\%  \\ 
USS reported FSC for pre-ret DR of gilts+3.0\% & 31.8\%  &  19.2\%  \\
\hline
FSC change for pre-ret DR increase of 0.5ppt & -2.7ppt & -1.4ppt  \\
(row 2 minus row 1) &  &  \\

\hline
FSC change for pre-ret DR increase of 2.7ppt & -15ppt & -8ppt    \\
(row 3 divided by 0.5 multiplied by 2.7) &  &    \\
\hline
\hline
Annual savings to sector from  FSC change &   &   \\
due to pre-ret DR increase of 2.7ppt & \pounds 1.5bn & \pounds 0.8bn    \\
(using 10ppt of FSC change equals \pounds 1bn) &  &   \\
\hline

\end{tabular}

\caption{USS reported sensitivity (row 1 and 2) of FSCs to changes in pre-retirement discount rate
from the 2020 and 2023 valuations \cite{USS_Val_TP_SI_2020,USS_Val_TP_SI_2023}. The subsequent rows then estimate the annual savings to the sector (from both 2020 and 2023 sensitivity) from FSC changes due to the 2.7ppt rise in the nominal pre-retirement DR seen between 2020 and 2023. The last line uses the known sector payroll, that is 100\%, as \pounds 10bn. }
    \label{tab:pre ret sensitivity}
\end{table}

Putting aside the fact that the difference between 2020 and 2023 USS sensitivity analysis suggests that the sensitivity analysis is itself volatile,
the reduction in FSCs due only to the change in the pre-ret DR can be estimated as 8ppt from the 2023 sensitivity and 15ppt from the 2020 sensitivity.  This corresponds to a net annual saving of \pounds 0.8bn (2023 sensitivity) or \pounds 1.5bn (2020 sensitivity) from only the changes to the pre-ret DR made between 2020 and 2023.

The total FSC reduction from 2020 to 2023 was 16.4ppt, falling from 37\% to 20.6\% as seen in Figure \ref{fig:gilt corr val 14-23}. It seems reasonable to use the 2023 sensitivity data of Table \ref{tab:pre ret sensitivity} to attribute 8ppt of the FSC reduction to the increase in the pre-ret DR. This means around half of the 16ppt reduction in FSCs between 2020 and 2023 can be attributed to the change in the pre-ret DR and half to the post-ret DR. 

This suggests that, despite their very different nature, both portfolios (respectively 90\% equities and 90\% bonds) are contributing equally to driving the almost perfect gilt yield dependence of the contribution rates. 

\newpage

\subsection{Technical Provisions gilt yield dependence}
\label{subsec:TP DR gilt yield}

Deficit Recovery Contributions are the cost per year of paying off any USS calculated Technical Provisions (TP) deficits. The TP deficit is TP liabilities minus assets.  
TP liabilities are the present value of cashflows (due to past promised benefits) discounted using a TP DR. As with the FSC DR, the TP DR is set using the Dual Discount Rate approach from the pre- and post-retirement discount rates.

Under the DDR approach USS set the TP DR as 33\%  pre-ret DR and 67\% post-ret DR in both 2020 and 2023\footnote{Weightings calculated from pp19-21 of USS 2023 TP consultation \cite{USS_Val_TP_SI_2023} details on github \cite{SussexUCUgithub}.}. So the TP DR is majority set by the post-retirement DR. 

\begin{figure}[hb!]

\centering

\includegraphics[width=0.9\textwidth]{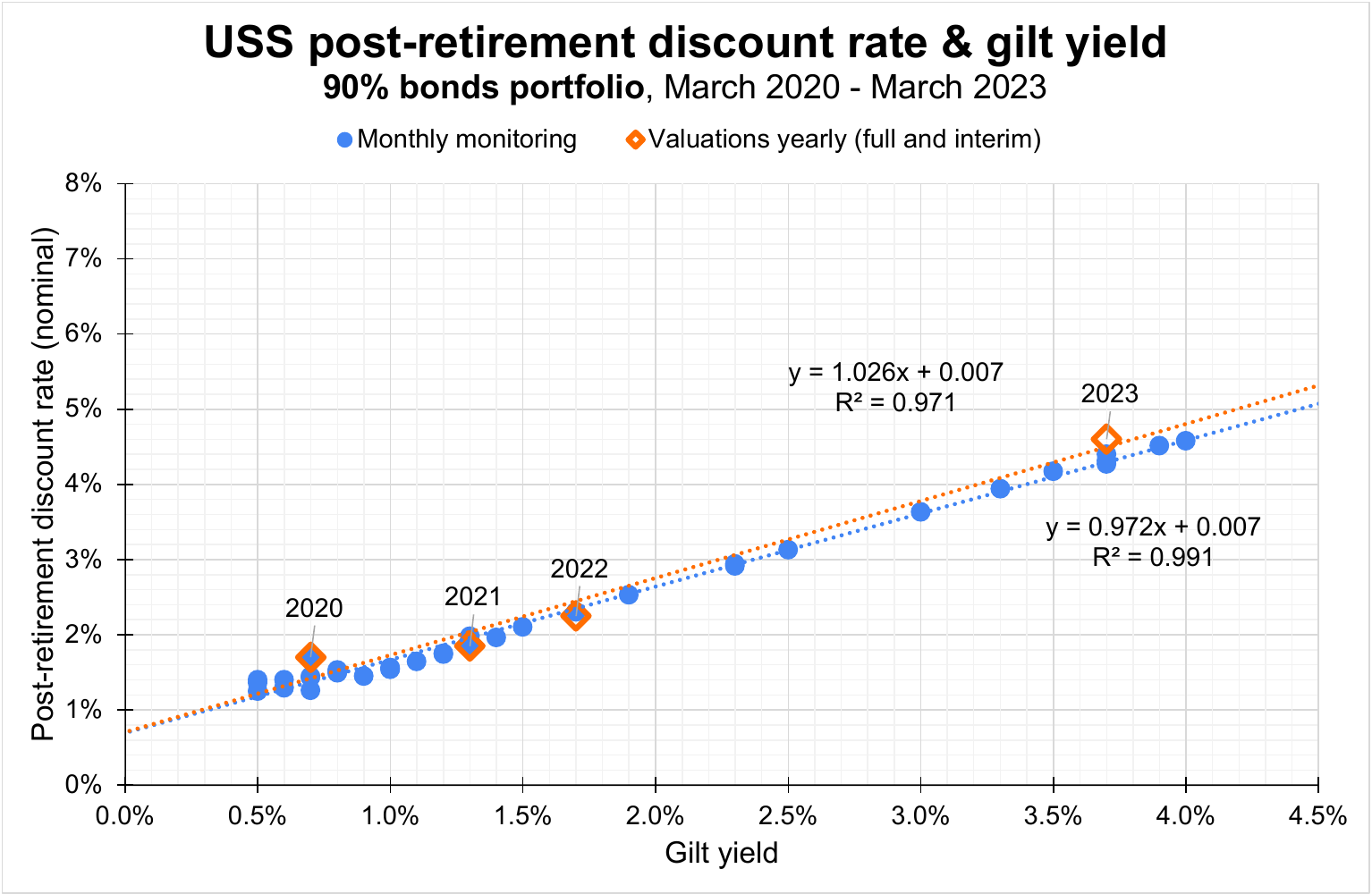}

  \caption{USS post-retirement discount rate which is the prudently adjusted annual expected return on a 90\% bonds portfolio. Monthly monitoring data since March 2020 are shown for annual full or interim valuations. The prudent returns show a 97\% and a 99\% correlation with the gilt yield.
  }
    \label{fig:post-ret_pru}

\end{figure}

The post-ret DR monitoring, March 2020-March 2022, is shown in Figure \ref{fig:post-ret_pru} to be 99\% correlated with the gilt yield. The valuation data is also shown for the four full and interim valuations over this period and demonstrates the same dependence.  It is worth noting that the gradient or slope is 1.0 for both sets of data, so the post-retirement DR is the gilt yield plus a bit.

As demonstrated in the previous section, the pre-ret DR shows a similarly very high correlation to the gilt yield  ($R^2$=97-99\% of Figure \ref{fig:gilt corr pre-ret pru 2020-2023}) as seen in the post-ret DR ($R^2$=97-99\% of Figure \ref{fig:post-ret_pru}).  It is less surprising that the post-ret DR (as compared with the pre-ret DR) has a correlation with the gilt yield, given that the post-ret portfolio is 90\% bonds. But it is still remarkable that the dependence is so high and to the extent that the \textit{post-ret DR is the gilt yield value of that month plus  around 0.7\%}. This result would not be expected if views on the long term nature of bonds were factored into decisions on the discount rates.

As an aside, it is worth noting that since the pre- and post-ret portfolios are respectively 90\% and 10\% equities (the remaining allocated as bonds), the TP DR appears to be calculated from an effective portfolio of  36\% equities and 64\% bonds. 
This is below the FSC effective equities allocation of 54\% (see Section \ref{sec:Pre-ret}), meaning both the  FSC and TP portfolios appear to be effectively below the Valuation Investment Strategy allocation of 60\% equities (see Glossary \ref{glossary}). 

Returning to a consideration of the DRCs. The TP DR is used to calculate the TP liabilities and hence the TP deficit. Once calculated, the total value of the TP deficit is divided by the recovery period (or number of years of deficit payments), and then divided by annual salary costs, to obtain a percentage of yearly payroll. So both DRCs and FSCs are calculated directly from pre- and post-ret discount rates and expressed as a percentage of salary or annual payroll.

USS reported a \pounds 15bn TP deficit in 2020, claiming this necessitated DRCs of 6.2\% of payroll 
for 18 years.  However, as the gilt yield moved upwards from  0.7\% in 2020 to 3.7\% in 2023 the USS estimated TP deficit rapidly transformed to a surplus of \pounds 7bn, rendering any DRCs unnecessary. 

DRC payments of around  \pounds 0.6bn a year were introduced in April 2022 from the 2020 valuation. Although formally paid by employers, the continuity of 70:30 split for \textit{all contributions}\footnote{This share of contributions is often confused with the default rule in the absence of an agreement, which is 65:35, see `contributions' and `cost-sharing' in Glossary \ref{glossary}.} meant employees effectively contributed.
The DRCs were halted in January 2024 as part of the agreement reached for the 2023 valuation.

The gilt yield dependence of USS valuation TP liabilities has been previously examined by Wong for the seven annual full and interim valuations 2011-2017. The USS TP liabilities showed a 99\% correlation with Index Linked 20-year Gilts \cite{Woon_CDF_2018/12}. The same data shows a 97\% correlation with Bank of England 20-year gilt yields. Using the approach by Wong, and using only USS data, the  TP liability dependence on gilt yield for 2020 monitoring is also shown to have a high correlation. This work is discussed in the Appendix \ref{app:gilt yield}.

\subsection{Summary of contribution rate dependence on gilt yield} \label{sec:summ_gilt}

In summary all available USS data from the last decade shows that contribution rates (which USS calculate using FSC and TP discount rates from pre- and post-retirement discount rates) are very highly dependent on the return on UK government bonds (the gilt yield) to the extent that the gilt yield can explain 95\% to 99\% of changes or volatility in the contribution rate. This gilt yield dependence is driving the volatility in costs through the gilt yield dependence of both the pre- and post-retirement discount rates.

\begin{table}[h]
    \centering
\begin{tabular}{|l|r|r|}
\hline
 \textbf{}& \multicolumn{2}{c|}{\textbf{Reduction 2020 to 2023  }} \\
 \textbf{Pre-April 2022 benefits}& \textbf{ppt salary } &  \textbf{\pounds bn / year}   \\
\hline
\hline
FSC reduction from 2020 to 2023  & 8ppt &  \pounds 0.8bn  \\
(due only to pre-ret DR change) & &    \\
\hline
FSC reduction from 2020 to 2023 & 8.4ppt & \pounds 0.8bn  \\
(due only to post-ret DR change) & &    \\
\hline
DRC reduction from 2020 to 2023 & 6.2ppt & \pounds 0.6bn    \\
(67\% post-, 33\% pre-ret DR change) & &    \\
\hline
\hline

Total reduction from 2020 to 2023 & 22.6ppt & \pounds 2.3bn    \\ 
\hline

\end{tabular}

\caption{The reduction in proposed contribution rates between 2020 and 2023 for maintaining pre-April 2022 benefits. Shown in percentage point (ppt) of salary and costs (\pounds bn/year) for USS pension scheme members and their employers. The reduction is separated between Future Service Costs and Deficit Recovery Contributions, and their dependence on pre- and post-retirement discount rates. }
    \label{tab:contributions_breakdown}
\end{table}

Table \ref{tab:contributions_breakdown} shows the change in costs between 2020 and 2023 that USS quoted to maintain a consistent level of benefits. The breakdown of costs seems to suggest that the pre- and post-retirement discount rate volatility are both contributing equally to the volatility in costs. The change in proposed costs are enormous, representing circa \pounds 2.3bn a year or equivalently 22.6\% of salary. 

This very high gilt yield dependence of the USS choice of discount rates (and hence costs) is to be contrasted with the much lower correlation, seen in USS data, between the gilt yield and the chosen assumptions for best estimates of returns equities. It seems that the judgements being made about assumptions for best estimates of returns are not feeding through to the discount rates.

The next sections considers how the USS valuation methodology explicitly drives the high gilt yield dependent volatility of contribution rates through the USS-specific definition of self-sufficiency.

\newpage

\section{Self-sufficiency liabilities are inflated}\label{sec:self sufficiency}

This section considers how USS modelling unnecessarily amplifies self-sufficiency (SfS) liabilities from the gilt yield and so unnecessarily increases contribution rates. Public statements by the Pensions Regulator on USS SfS liabilities, considered in Section \ref{sec:IRMF}, suggest that high estimates of SfS liabilities invite regulatory intervention due to the enlarged shortfall of assets relative to the high SfS liabilities. 

Self-sufficiency (SfS) was first introduced by USS in 2014. USS refers to SfS as their ‘primary basis and benchmark’ for measuring risk. In 2014 and 2017 SfS was defined within the larger mechanism of USS's notorious Test 1 \cite{Marsh_2018}.  USS replaced Test 1 in 2018\footnote{USS dropped Test 1 for the 2018  valuation following widespread criticism \cite{Marsh_2018}.}, but the USS self-sufficiency definition has remained. The 2020 USS Technical Provisions Consultation described SfS as follows\footnote{SfS has been defined by USS in various ways since 2014, as detailed in the Appendix \ref{app: self-sufficiency definitions}.}:

\begin{quote}
    
The [USS] Discussion Document (pages 21-22) proposed that self-sufficiency remains the primary basis, or benchmark, for measuring and managing funding risk in the Scheme. Self-sufficiency is a low-risk strategy for funding the Scheme in the absence of a covenant. \textit{ It corresponds to a confidence level of 95\% (equivalent to a 5\% failure rate) of being able to pay all benefits when they fall due without the need for any additional contributions,} \textbf{\textit{while maintaining a high funding ratio.}}  [Added emphasis italic and bold. USS 2020 TP Consultation \cite{USS_Val_TP_SI_2020}.]
\end{quote}

The above definition includes two distinct conditions. For ease of discussion these will be referred to as  the ‘benefit payment' and the ‘funding ratio'  conditions. These two conditions are explored in a key USS 2018 Response document \cite{Otsuka_response_2018} which states that the funding ratio condition is consistent with the parameters of Test 1 and so the funding ratio condition may have been in place since 2014.

In 2023 USS stated that the required funding ratio is set to 90\%, which is the same value used in the USS 2018 Response. In addition, the SfS funding ratio was, for the first time, explicitly stated to be required at a 95\% confidence level every three years. So the two SfS conditions can now be separated and clarified as:

\begin{itemize}
\item [1.] A confidence level of 95\% (equivalent to a 5\% failure rate) of being able to pay all benefits when they fall due without the need for any additional contributions [the benefit payment condition], and
\item [2.] A confidence level of 95\% (equivalent to a 5\% failure rate) of achieving a 90\% funding ratio every three years [the funding ratio condition].

\end{itemize}

The SfS conditions are considered in detail in Sections \ref{subsec:SfS costs} to \ref{subsec:sfs_independent_analysis}. In short: USS choose a SfS investment strategy (the SfS portfolio) which is set at 10\% equities and 90\% bonds from day one, even though USS report that the portfolio would take a decade to implement \cite{USS_Transition_Risk_2022}. USS applies both benefit payment and funding ratio conditions to this SfS portfolio in run-off (with no contributions into the scheme) and without covenant support (additional contributions from employers)\footnote{Run-off and covenant support are defined further in the Glossary \ref{glossary}.}.

The funding ratio condition is shown in general to dominate over the benefit payment condition, and by a significant margin. This strongly indicates that the funding ratio condition is setting the SfS liabilities\footnote{USS indicate the funding ratio also dominates the 2023 valuation, Sec. \ref{sec:SfS_fund_ratio} and App. \ref{app:USS sfs 2023}.}.   
The funding ratio condition is also shown to dominate at relatively early times, and this is also supported by USS's 2023 statements. 
Figure \ref{fig:Coughlan} shows that (for the only available USS data) 85\% of cases that failed the funding ratio condition in year three go on to pay benefits in full. So the SfS liabilities are set from the early time behaviour of a funding ratio condition that does not predict future failure to pay pensions. The condition is applied from day one to a SfS portfolio that would take a decade to implement.

\subsection{USS self-sufficiency sets costs}\label{subsec:SfS costs}

In the USS methodology the SfS liabilities are critically important in determining contribution rates. As discussed in Section \ref{sec:Pre-ret} contribution rates consist of two parts: the Future Service Costs (FSCs) and the Deficit Recovery Contributions (DRCs). Since 2020 USS has used a Dual Discount Rate (DDR) of pre- and post-retirement discount rates (pre- and post-ret DRs) to calculate FSCs and DRCs. For the 2020 valuation and monitoring USS explicitly set the post-ret DR equal to the SfS DR. For the 2023 valuation the post-ret DR was uncoupled from the SfS DR and currently sits a little above (0.3-0.4ppt) the SfS DR. As remarked by AON, changes to SfS CPI assumptions suggest that the `overall effect is to maintain a similar level of prudence [from 2020 to 2023] in the self-sufficiency approach relative to the technical provisions post-retirement liabilities' \cite{AON_2023}.

Once the SfS liabilities are set, then the SfS DR and post-ret DR are set. The FSC DR is 45\% post-ret DR and the TP DR is 67\% post-ret DR\footnote{USS provide no justification for these ratios. Calculated from pp19-21 of USS 2023 TP \cite{USS_Val_TP_SI_2023}.}. 
Section \ref{sec:Tar_rel} considers how the USS metrics tether the TP liabilities to the SfS liabilities and so set the TP DR. This in turn sets the pre-ret DR. Hence both the pre- and post-ret DRs are set from the SfS liabilities, which set the contribution rates via FSCs and DRCs. So the SfS liabilities are critically important to costs - it really matters how they are calculated. The next four subsections explore the USS SfS calculation in detail using USS data and preliminary independent analysis. 

\subsection{USS SfS only uses the funding ratio condition}\label{subsec:SfS only funding ratio}

\begin{figure}[hb!]

\centering

\includegraphics[width=0.9\textwidth]{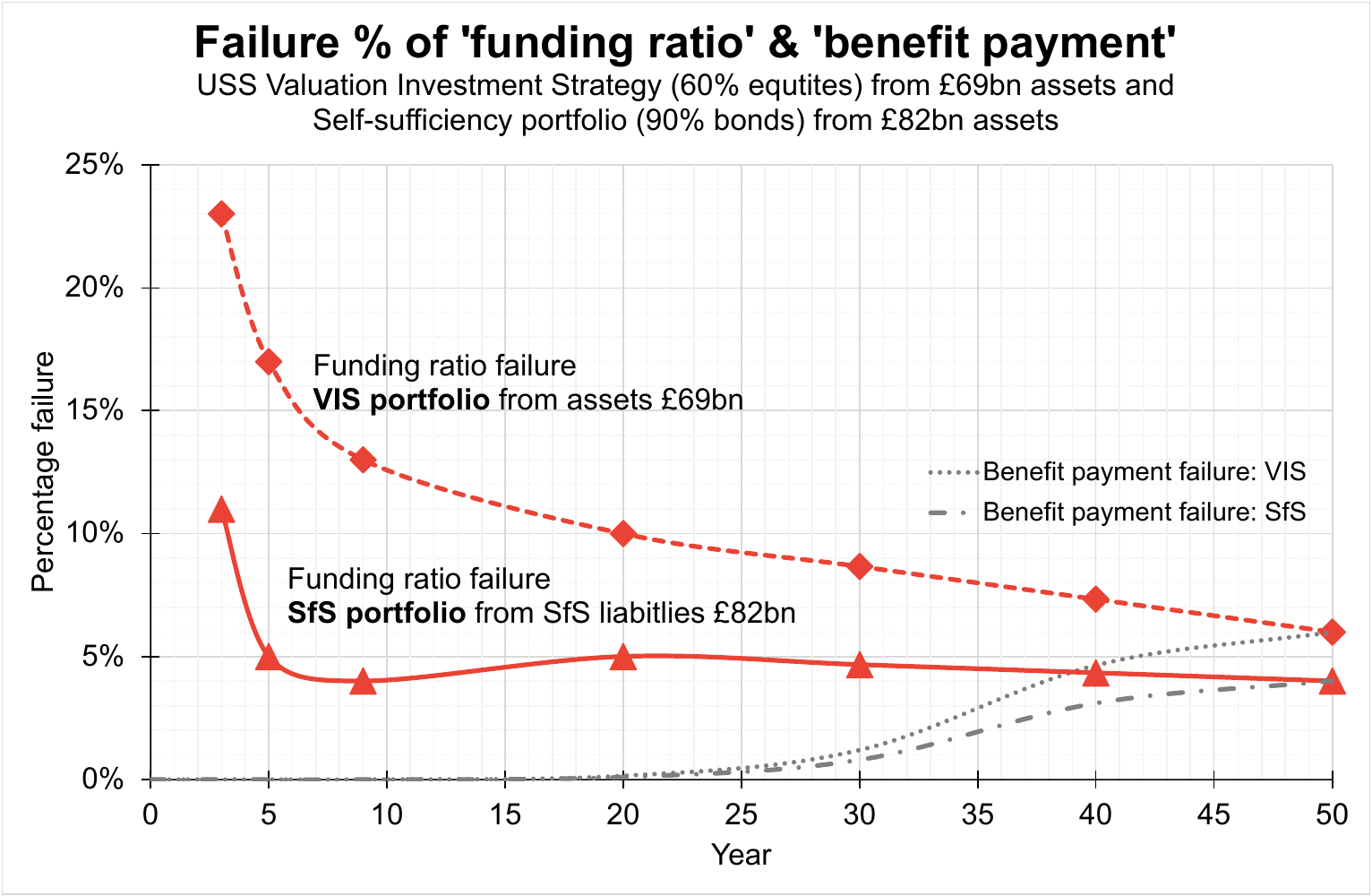}

  \caption{USS 2018 data showing probability of failing the 90\% funding ratio condition (red) and the benefit payment condition (grey). The conditions are applied to two USS portfolios: the Valuation Investment Strategy or Reference portfolio (60\% equities) from assets of \pounds 69bn and the SfS portfolio (90\% bonds) with initial assets of \pounds 82bn. USS assumptions on returns are detailed  on the last page of \cite{Otsuka_response_2018}. 
  USS data have been smoothly interpolated.
  }
    \label{fig:2018_SfS_VIS}

\end{figure}

There is little public analysis on the USS SfS definition and almost nothing on the funding ratio. However, general properties can be understood from two USS documents and some independent analysis. First, the 2018 USS Response is believed to contain the only data on the two separate SfS conditions \cite{Otsuka_response_2018}. Second, the USS 2023 TP Consultation discusses sensitivity of the funding ratio condition, see Appendix \ref{app:USS sfs 2023}. Independent analysis is included in Section \ref{subsec:sfs_independent_analysis}.

Figure \ref{fig:2018_SfS_VIS} shows the percentage of paths that fail the funding ratio condition for two USS portfolios: the Valuation Investment Strategy (VIS) of around 60\% equities and 40\% bonds with initial assets at \pounds69bn. The VIS portfolio is a close approximation to the actual USS investment strategy. The second is the SfS portfolio of 90\% bonds with initial assets \pounds82bn. This value of the SfS initial assets was selected by USS as initial assets of the VIS portfolio at \pounds69bn plus \pounds13bn, the present value of `covenant support' set at 7\% of payroll for 20 years. 

In the USS simulations of 2018, the funding ratio is measured every year as assets divided by SfS liabilities. SfS liabilities are calculated from remaining cashflows of promised benefits using a SfS discount rate. So the paths of the simulation fail the 90\% funding ratio condition in a given year if the assets drop below 90\% of the remaining SfS liabilities. It is worth noting that the mechanism to calculate the SfS DR from the SfS liabilities requires a SfS DR as an input. 

It is also important to understand that the funding ratio failure shown in Figure \ref{fig:2018_SfS_VIS} is recorded as failure percentage \textit{in each year}, but the benefit payment failure is recorded \textit{cumulatively over time}. This is because a path can fail the 90\% funding ratio condition multiple times (as the assets increase or decrease over time relative to the remaining SfS liabilities) and so a cumulative accounting would double, triple count etc. However a path can only fail a benefit payment condition once, when the scheme has actually run out of money, so can usefully be cumulated. 

A first point to note about the funding condition failure shown in Figure \ref{fig:2018_SfS_VIS} is that it is dominated by early-time behaviour and then decreases rapidly over time. So for a SfS portfolio of 90\% bonds the condition will be driven by short-term (first decade) gilt yield values. This is to be compared with the pension payment condition (shown in grey for the two portfolios) which is negligible in the early years and only becomes significant at later decades, when liabilities are lower.

A second point to notice is the difference between the failure rates of the funding ratio condition (red) and the pension payment condition (grey). The funding ratio condition failure rate \textit{per year} in year three is at least 3 to 4 times larger (24\% and 11\% and for VIS and SfS respectively) than the \textit{total cumulative} benefit payment failure rate (6\% and 4\%) at year fifty. 

These first two points combine to demonstrate an early-time dominance of the funding ratio condition in setting the SfS liabilities. The next section explores the relevance of this by considering Figure \ref{fig:Coughlan} in detail. It is also noted again that the SfS portfolio would, by USS's own estimates take a decade to implement.

A final point to observe is that the funding ratio and the benefit payment conditions asymptote to the same value. This is because once all benefits have been paid any path that has run out of money (has no assets) will automatically fail the funding ratio condition each year. Similarly paths that have assets remaining but no benefits left to pay will automatically pass the funding ratio condition in that year. So benefit payment failure and funding ratio failure do asymptote to the same value, but only toward the limit when all benefits have been paid. 

The suitability of the funding ratio condition to measure ability to pay pensions is explored in the next section and also in the independent analysis in Section \ref{subsec:sfs_independent_analysis}.

\subsection{The SfS funding ratio condition is not useful}\label{sec:SfS_fund_ratio}

\begin{figure}[hb!]
\centering

\includegraphics[width=0.9\textwidth]{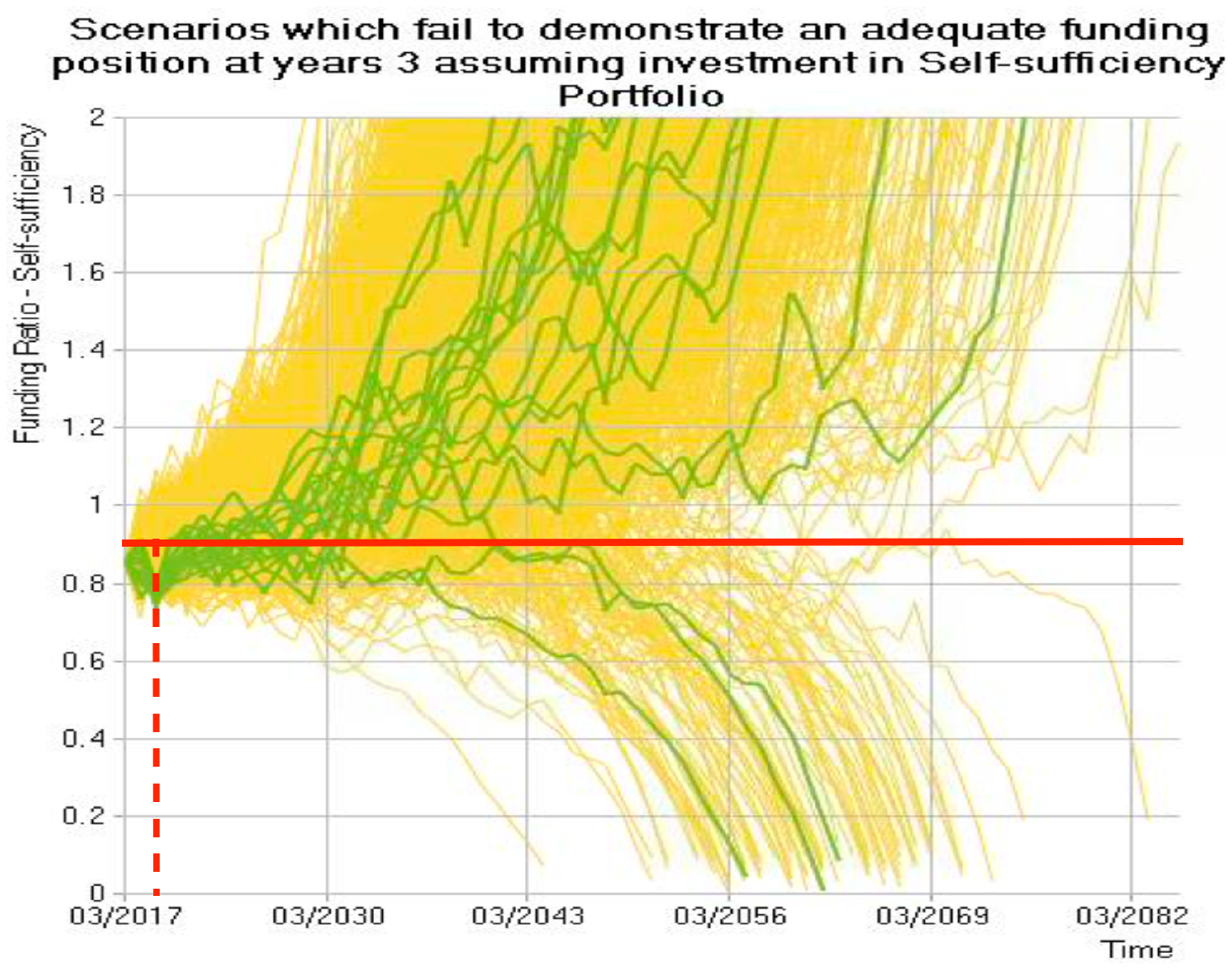}

  \caption{A graph from the USS 2018 Response paper shows a SfS portfolio with initial assets of  \pounds 82 billion. The circa 2,000  yellow lines represent all paths in the stochastic simulation. The thicker green lines represent the 11\% of paths that failed the 90\% funding ratio condition at year 3. The red horizontal line marking the 90\% level and the red dashed vertical line marking year three have been added. USS assumptions on returns are detailed  on the last page of \cite{Otsuka_response_2018}.
 }
    \label{fig:Coughlan}

\end{figure}

This section explores the funding ratio condition by considering Figure \ref{fig:Coughlan}.
and leads to the conclusion that the funding ratio condition is not useful in determining if benefits can be paid, except toward the limit when benefits have been paid.

Figure \ref{fig:Coughlan} shows USS data for around 2,000 simulations of a SfS portfolio from initial assets of \pounds82bn. The \pounds82bn is calculated as the actual assets of \pounds69bn plus the \pounds 13bn representing the covenant support employers could provide discounted to a net present value. This was estimated at the time for this simulation as 7\% of payroll for 20 years. The simulation uses input parameters for expected returns on the investment strategy as detailed in the USS paper \cite{Otsuka_response_2018}. The simulation models a scheme in run-off, so paying promised benefits only from the assets and investment returns. The yellow paths represent a distribution of possible outcomes as the assets grow (or decrease) while paying promised benefits as they fall due.

The principle of carrying out such a simulation could be to try to answer the question: What is the likelihood of being able to pay promised benefits, as they fall due, from the current assets? This is a reasonable question - but it is important to check how the simulation is attempting to answer such a question. The yellow lines show all paths of the simulation and the horizontal axis shows the 65 years of the simulation, from the 2017 to 2082. The vertical axis shows the funding ratio for each path as measured in each year of the simulation. 

The dark green, thicker lines represent those paths (among all 2,000 yellow paths) that have failed the funding ratio condition in year three. Failing the funding ratio condition means that the funding ratio is less than 90\%, so assets are less than 90\% of the SfS liabilities. Added to the image is a horizontal red line of this 90\% boundary and a red vertical dashed line marking year three. As expected the dark green paths are those that drop below 90\% at year three. 

As the yellow and green paths evolve in time they either move upwards or downwards. If a path moves up the assets are growing faster than pensions need to be paid. If a path moves down then its assets are dropping, it may recover and move up, but if it is moving above 100\% at the end of the simulation it has succeeded in paying all pensions in run-off from a SfS portfolio with covenant support added, as a net present value, at the beginning.

The fundamental point to note in Figure \ref{fig:Coughlan} is that of the circa 20 green paths that represent failure of the funding ratio condition at year three, around 17 of them go on to improve their funding position and pay pensions in full. The remaining three that stay below the 90\% funding ratio line are unlikely to pay pensions, but it is not clear that they are unable. 
In summary, Figure \ref{fig:Coughlan} tells us that for 2018 data at least 85\% of paths that fail the SfS funding ratio condition in year three go on to pay all pensions in full. 

Turning next to the USS 2023 discussion of the funding ratio condition as reproduced in Appendix \ref{app:USS sfs 2023}. USS makes clear for the first time in any consultation material that their SfS modelling `comfortably passes' the benefit payment condition but that the funding ratio condition is not quite passed. Any reasonable reading would suggest that the funding ratio condition dominates and therefore sets the SfS liabilities for the 2023 valuation. The same USS discussion also makes clear that the funding ratio is highly sensitive to the input assumptions and particularly binding in the early years of the simulation.  

It is clear that the funding ratio condition is not a useful test to impose in the early years of a simulation if the question is whether the pensions can be paid. Yet it is the condition imposed by USS that appears to be responsible for setting the SfS liabilities and producing high sensitivity to the gilt yield behaviour.

\subsection{The benefit payment condition}\label{subsec:benefit payment}

The funding ratio condition as described in Section \ref{sec:SfS_fund_ratio} does not measure the ability to pay pension benefits. It is also clear from  Section \ref{subsec:SfS only funding ratio} that the funding ratio condition dominates in setting the SfS liabilities. This means the funding ratio condition obscures the other SfS condition, the benefit payment condition.  

This section considers the benefit payment condition, as introduced in Section \ref{sec:self sufficiency}, which attempts to answer the question: How likely is it that USS can pay all promised pensions, as they fall due, from the assets? This is a question that aligns with the statutory funding objective of the UK Pensions Act 2004 \cite{UKParl_PensionsAct_2005_S2222}: 
\begin{enumerate}
    \item [(1)]Every scheme is subject to a requirement (“the statutory funding objective”) that it must have sufficient and appropriate assets to cover its technical provisions.
    \item [(2)]A scheme’s “technical provisions”means the amount required, on an actuarial calculation, to make provision for the scheme’s liabilities.
\end{enumerate}

\begin{figure}[hb!]

\centering

\includegraphics[width=0.9\textwidth]{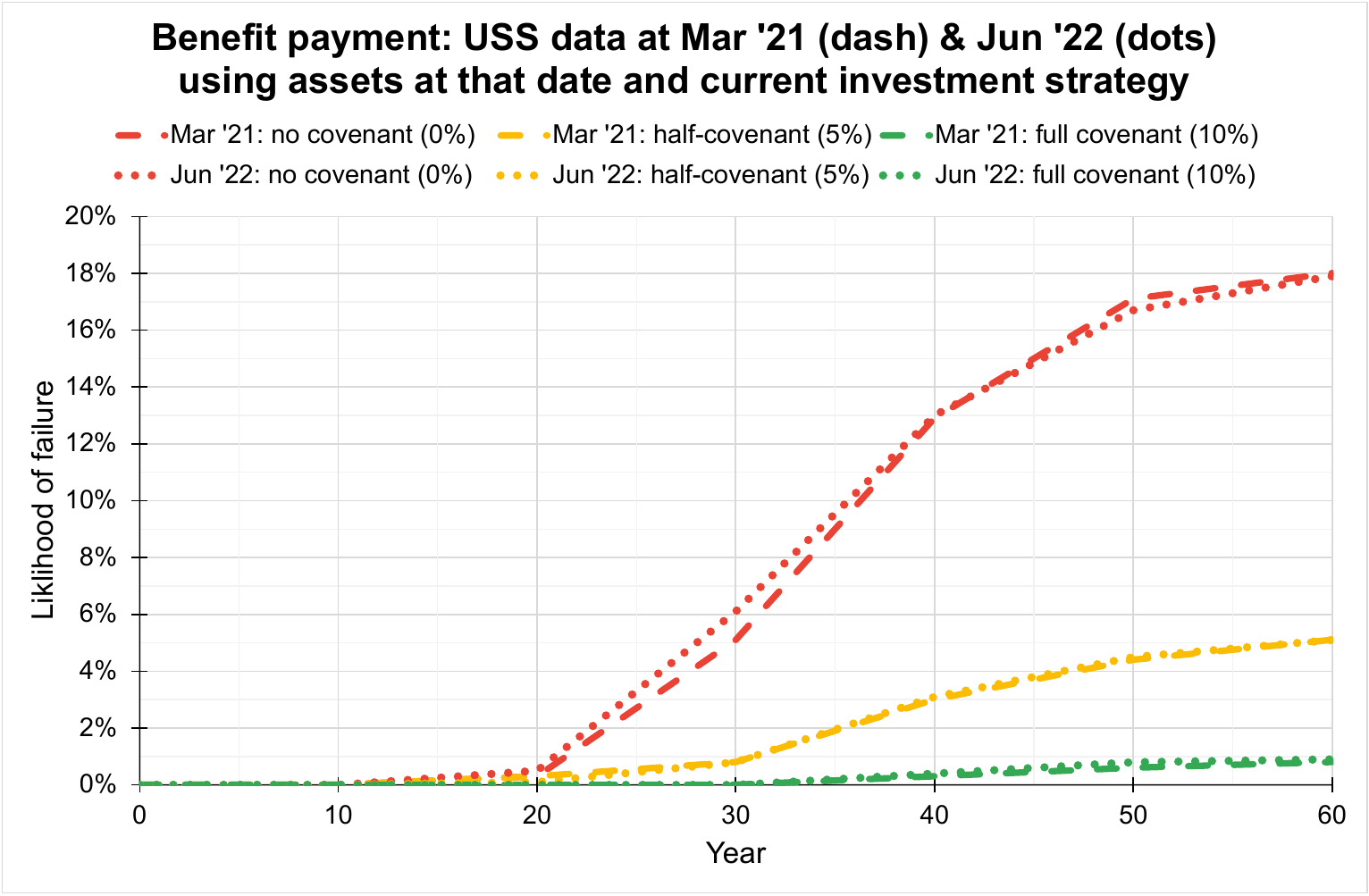}
  \caption{USS data on benefit payment (or capital exhaustion) condition failure rates from the current investment strategy at March 2021 (dashes) and June 2022 (dots) \cite{USS_Briefings_Analysis}. The results appear highly stable, that is unchanged, against the significant changes in USS assumptions between March 2021 and June 2022.
 }
    \label{fig:Benefit_payment}

\end{figure}

There are three main sources of analysis of the benefit payment condition\footnote{The benefit payment condition is referred to by USS as the capital exhaustion condition.} in the context of USS \cite{Miles_Sefton_2021, davies2021uss, USS_Briefings_Analysis} and some criticisms \cite{Con-capital_exhaustion} . All three adopt the same approach of simulating the scheme assets, from some fixed date, as they evolve in time, while paying all benefits promised up to that date, from the assets and returns on assets. All three pieces of analysis produce broadly similar results that usefully demonstrate the sensitivity of the model to assumptions about investment returns and portfolio weighting. The sensitivity to portfolio weighting, as shown in Figure 2 of \cite{davies2021uss}, is particularly striking as it demonstrates that portfolios with a higher percentage of bonds are much more likely to run out of funds. 

Of the three sources of analysis, only the USS approach also includes modelling of financial support from the employer in the form of the covenant.  Figure \ref{fig:Benefit_payment} shows data from this USS analysis, applied to the USS current investment strategy of 60\% equities, at the two dates of March 2021 (solid line) and June 2022 (dots). USS considers three different scenarios at each date depending on the level of support from employers. The simulation with no covenant support from employers shows a cumulative likelihood of 18\% of failing to pay all benefits. If covenant is added at 5\% of payroll for 30 years, the likelihood of failing is reduced to 5\%. If full covenant of 10\% over 30 years is added the likelihood of paying all benefits is 99\%. 

It is important to note that the two profiles of benefit payment failure are almost identical for March 2021 and June 2022. Table \ref{tab: benefit payment} shows how different the USS costs, surpluses and metrics were between these dates, due (as shown in Section \ref{sec:gilt_yield}) almost exclusively to changes in the gilt yield change. However the benefit payment condition of the SfS definition is remarkable for the level of stability it demonstrates between these two dates. This level of stability of the benefit payment condition is not reflected in any of the USS metrics or costs.

\begin{table}[!ht]
    \centering
    \begin{tabular}{|r||r|r|}
    \hline
       & March 2021 & June 2022 \\ \hline \hline
        USS gilt yield & 1.30\% & 2.30\% \\ \hline
        Future Service Costs & 36.70\% & 27.40\% \\ \hline
           Assets & \pounds  80bn & \pounds  80bn \\ \hline
        TP surplus & - \pounds  7.6bn & - \pounds 1.8bn \\ \hline
        SS surplus & - \pounds  31.2bn & - \pounds  13.6bn \\ \hline
        Metrics  & Mostly Red & All Green \\ \hline
    \end{tabular}
    \caption{The gilt yield, future service costs, assets, TP and SfS surplus at March 2021 and June 2022. With the exception of the assets, the USS data including costs is very different between the two dates. This is to be contrasted with the benefit payment condition profiles of Figure \ref{fig:Benefit_payment}. All data from USS monitoring \cite{USS_vals_mon}.  } \label{tab: benefit payment}
\end{table}

\newpage

\subsection{An independent analysis of USS self-sufficiency} \label{subsec:sfs_independent_analysis}

Preliminary independent analysis of both USS self-sufficiency conditions is now considered. This work shows agreement with the observations drawn so far from analysis of USS data:
that is the funding ratio condition dominates in setting the SfS liabilities (and so inflates their value relative to their value from only the benefit payment condition) and secondly, that failure of the funding ratio condition does not predict failure to pay benefits.

\begin{figure}[hb!]

\centering

\includegraphics[width=0.48\textwidth]{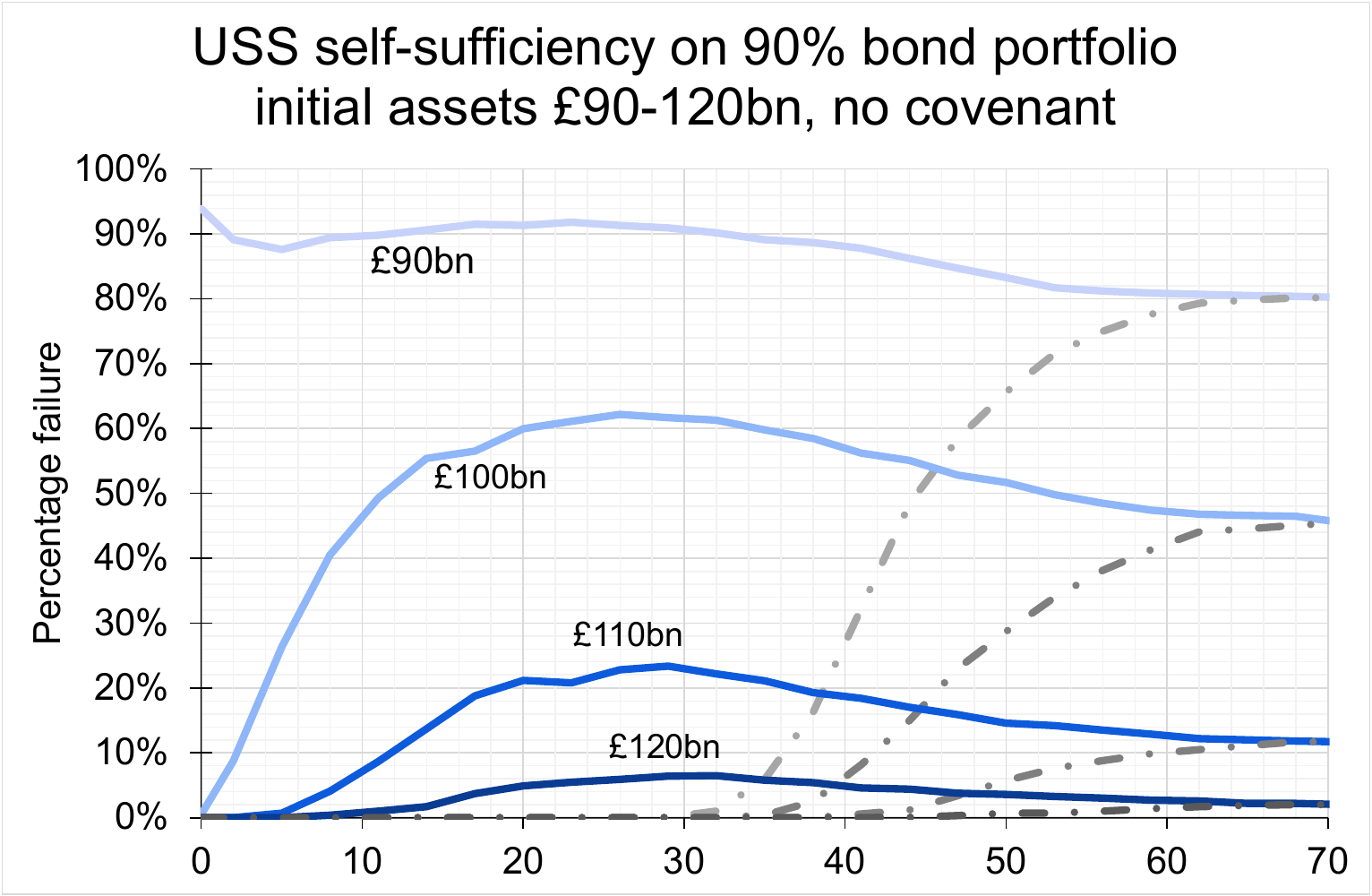}
\includegraphics[width=0.48\textwidth]{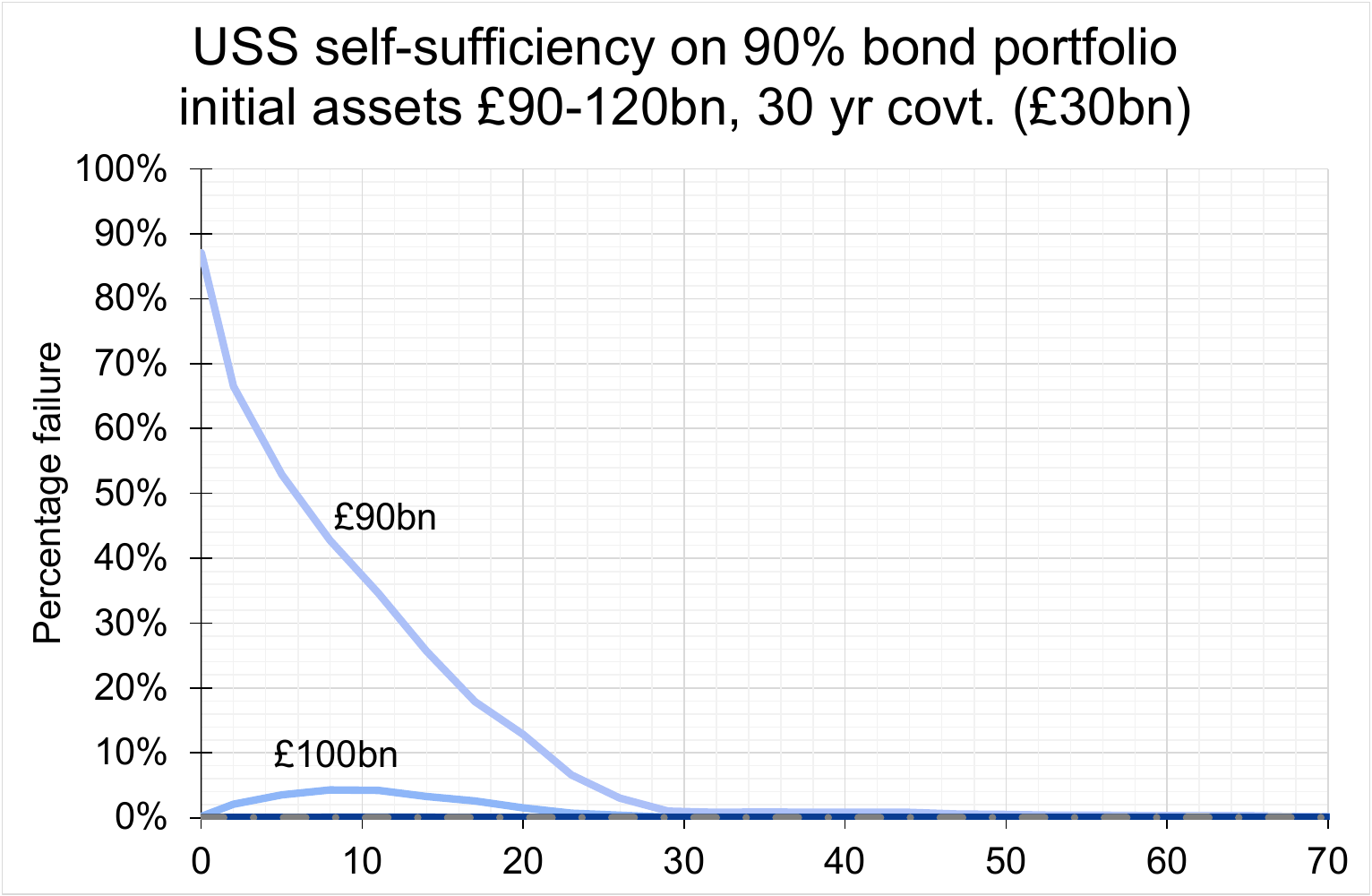}
\includegraphics[width=0.48\textwidth]{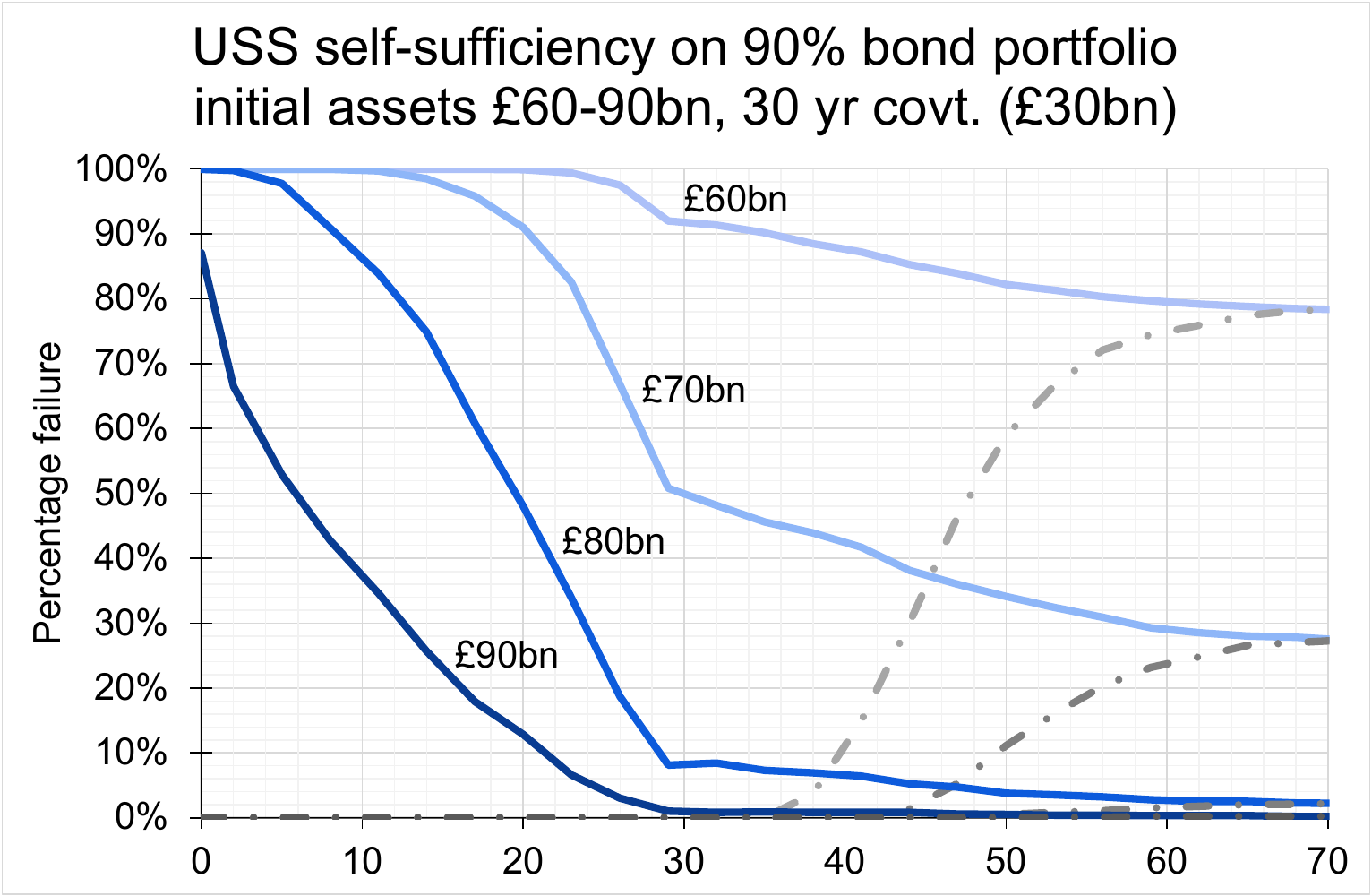}
\includegraphics[width=0.48\textwidth]{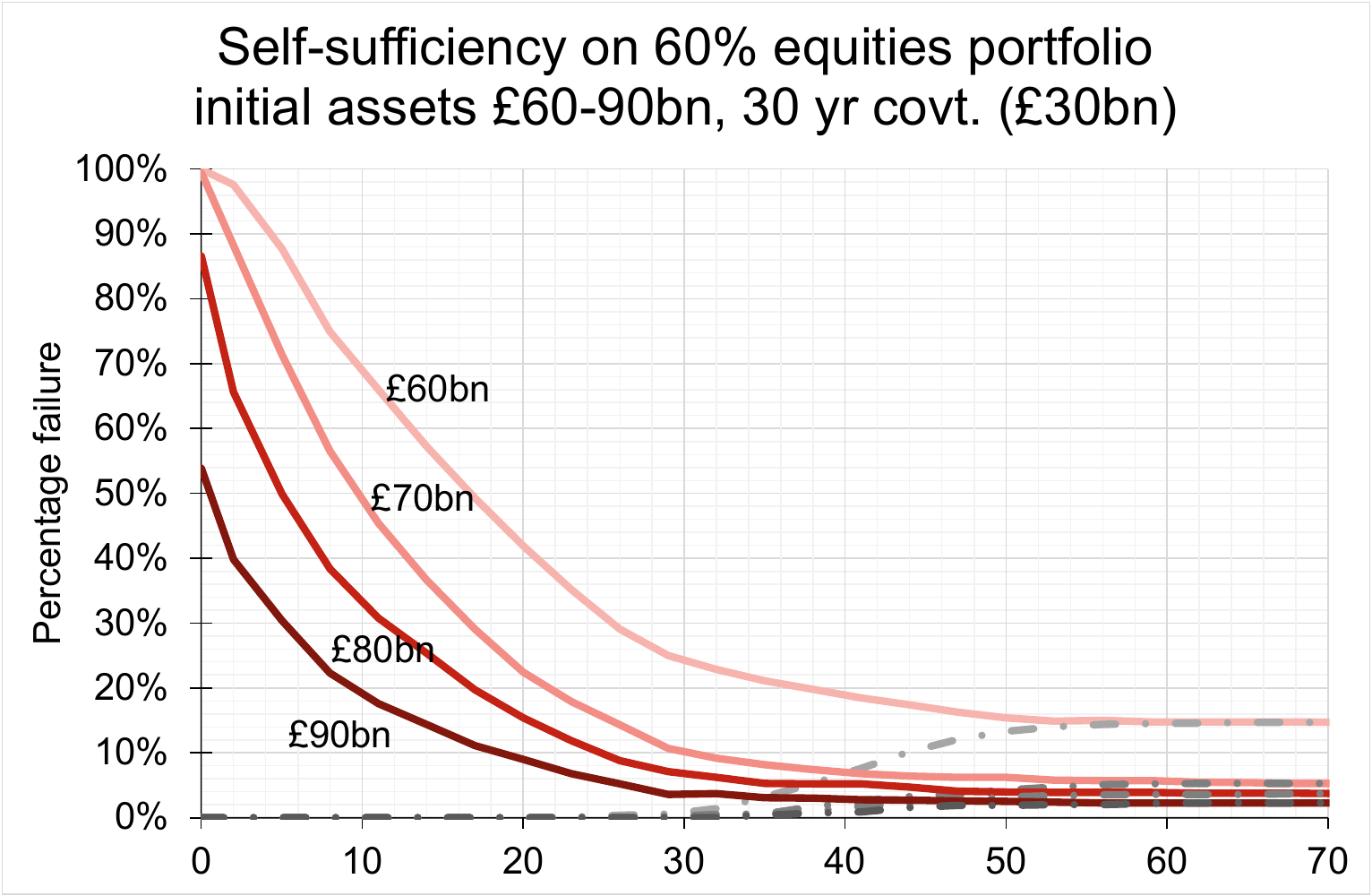}

\caption{Preliminary independent analysis of the behaviour of USS-style self-sufficiency conditions. All data and analysis available at \cite{SussexUCUgithub}. The failure rates for funding ratio (solid lines) and the benefit payment (dot-dash) are shown for four situations. The modelling follows Miles and Sefton \cite{Miles_Sefton_2021} with  mean equity return 4.5\%, standard deviation 17.5\%, mean bond return -1.0\%, standard deviation 2.0\%. 1000 runs, rebalanced portfolio with no mean reversion. The SfS funding ratio condition uses input discount rate of -0.75\%, equivalent to SfS net present cashflows of $\pounds$100bn. Top left: 90\% bonds, initial assets $\pounds$90-120bn. Top right: 90\% bonds, initial assets $\pounds$90-120bn with added covenant support (at 10\% of payroll for 30 years, amounting to circa \pounds 30bn). Bottom left: 90\% bonds, initial assets $\pounds$60-90bn with covenant support. Bottom right:  60\% equities, initial assets $\pounds$60-90bn with covenant support. The SfS requirement of $<$5\% failure for both conditions is not satisfied for any of the data top left or bottom left and right. 
} 
    \label{fig:SfS_indABCD}

\end{figure}

Figure \ref{fig:SfS_indABCD} shows the results of four sets of simulations following the approach of Miles and Sefton \cite{Miles_Sefton_2021, davies2021uss}. Failure rate for both the USS funding ratio condition (solid lines) and benefit payment condition (dot-dash) are applied from initial assets ranging from \pounds 60bn to \pounds 120bn as shown. Following the capital exhaustion analysis of USS \cite{USS_Briefings_Analysis}, covenant support from employers is also added to the simulations for three of the figures. This support takes the form of the USS employers' covenant, currently valued at 10\% of payroll for 30 years, added to each year of the simulation for the first 30 years. This has a present value of around $\pounds$30bn.

It is clear that the two conditions of the SfS definition have very different behaviour to each other for all the simulations shown. The analysis presented shows some of the same features as the USS data in Figure \ref{fig:2018_SfS_VIS}. The funding ratio condition is, like the USS data, an earlier time feature than the benefit payment condition. The funding ratio condition always dominates and the two conditions always asymptote to the same value as described in Section \ref{subsec:SfS only funding ratio}. As with the USS data, described in the same section, the funding ratio condition is shown per year, and the benefit payment condition shown cumulatively.

The high sensitivity of the funding ratio condition to the initial assets is evident in all four graphs. The sensitivity to  weighting of the portfolio towards bonds or equities is also evident by comparing the lower two figures, which use identical assumptions except that the figure on the right is 60\% equities and on the left is 90\% bonds. Further preliminary investigations suggest that other highly sensitive assumptions include the chosen value of the return on bonds, and the chosen value of the funding ratio test (set at 90\% for the figures shown as per USS's choice).

None of the simulations shown top left or bottom right and left, with initial assets $\pounds$90-120bn  satisfy \textit{both} of the USS SfS conditions at $>5\%$ confidence level. The figure on the top left most resembles the USS approach to calculating self-sufficiency and although the shape of the funding condition profiles vary\footnote{The simulated data either show an initial fall or a rise and fall, while the USS data show early time fall. In these preliminary investigations the early years profile of the funding ratio condition demonstrated very high sensitivity to early years profile of the projected cashflows and choice of SfS funding ratio. For example, if the SfS funding ratio condition was set as 110\% of TP liabilities the profiles were much closer to those of Figure \ref{fig:2018_SfS_VIS}. Further analysis is needed beyond this preliminary exploration to understand the behaviour of funding ratio profiles.}, they are not inconsistent with the USS data of Figure \ref{fig:2018_SfS_VIS}. This portfolio requires initial assets, or self-sufficiency liabilities of $\pounds$130bn to satisfy the dominant funding ratio condition, this is within the range that USS were reporting in the monitoring immediately after the 2020 valuation date. The figure on the bottom right most resembles USS's current investment strategy, and SfS liabilities of  $\pounds$65bn are required to pass the benefit payment condition at the 5\% level, while SfS liabilities of $\pounds$115bn are needed to pass the funding ratio condition to the same degree.

\begin{figure}[ht!]
\centering
\includegraphics[width=0.9\textwidth]{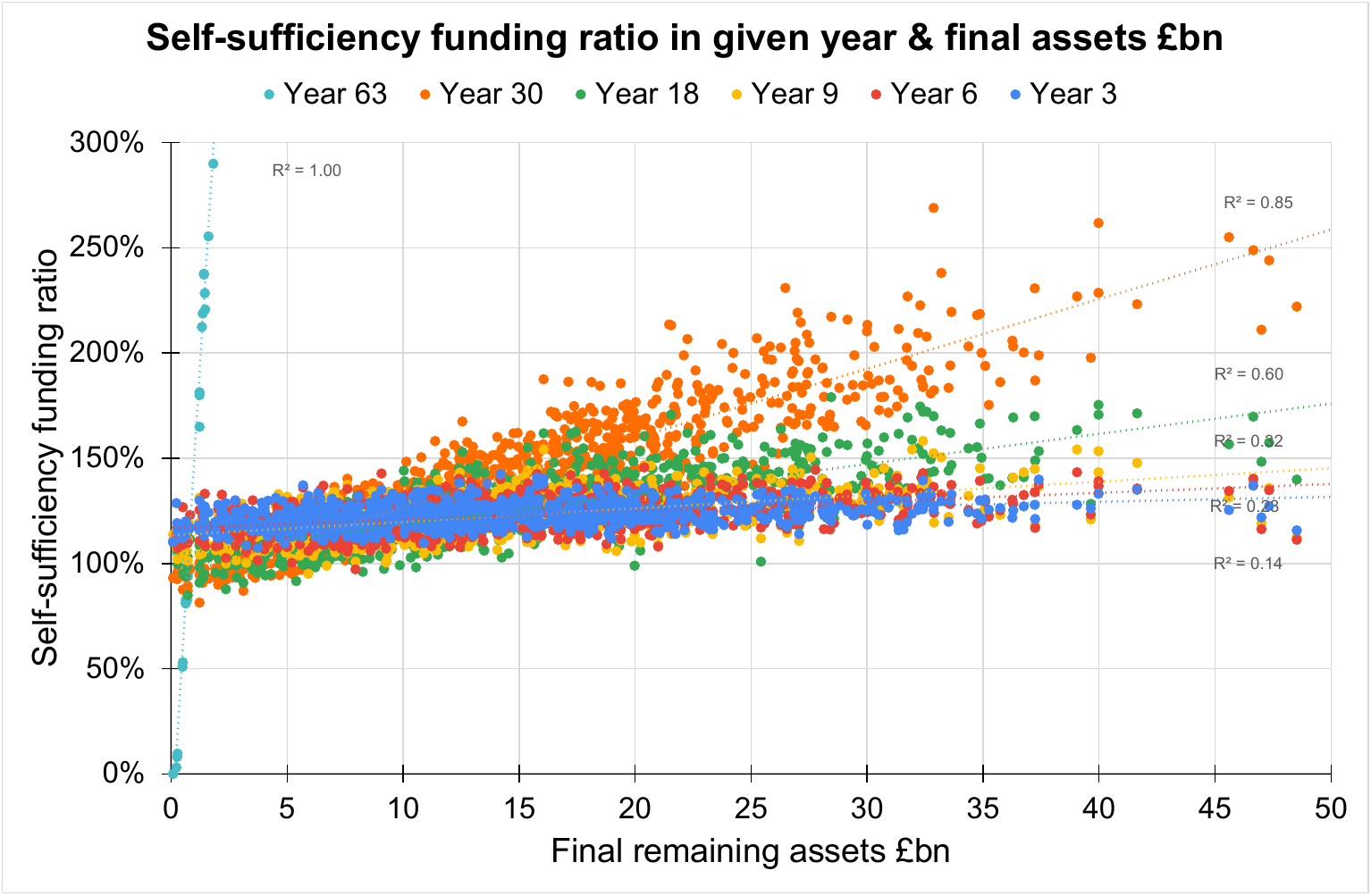}

  \caption{Self-sufficiency funding ratio (\%) at years 3, 6, 9, 18, 30 and 63 against final value of assets. Simulation data as top left of Figure \ref{fig:SfS_indABCD} with assets $\pounds$100bn.
  }
   \label{fig:FRC_correlation}
\end{figure}

Next, the ability of the funding ratio to predict scheme outcomes is considered. Figure \ref{fig:FRC_correlation} presents all data points for the funding ratio in years 3, 6, 9, 18, 30 and 63 against the final assets of the scheme for that path. 
Correlations are reproduced in Table \ref{tab:FRC_correlation}. The very low correlations in the early years make clear that the funding ratio condition is unable to predict the success of the scheme, where success is measured by the final value of the assets in $\pounds$bn after all benefits have been paid.   
  \begin{table}[!ht]
    \centering
    \begin{tabular}{|l|l|l|l|l|l|l|}
    \hline
        Year & 3 & 6 & 9 & 18 & 30 & 63 \\ \hline
        Correlation $R^2$ & 14\% & 23\% & 32\% & 60\% & 85\% & 100\% \\ \hline
    \end{tabular}
       \caption{Correlations of Figure \ref{fig:FRC_correlation}, between self-sufficiency funding ratio in a given year
and scheme success as measured by final assets $\pounds$bn.
  }
    \label{tab:FRC_correlation}
\end{table}

All the independent analysis shown here is preliminary, so all results should be caveated in this context. However, results suggest that the funding ratio condition always dominates and is therefore responsible for setting the SfS liabilities. Results also suggest that the funding ratio is not useful for predicting the long term behaviour of the scheme in terms of remaining assets  and so unlikely to predict failure. It is difficult to reconcile the importance of the funding ratio with the absence of available USS analysis on the properties of this self-sufficiency condition.

\newpage

\section{USS Actual \& Target Reliance metrics} \label{sec:ActTarRel}

Reliance is viewed by USS as the support available, in terms of money, from employers if the scheme enters financial difficulty. To try to quantify reliance, USS defines the Affordable Risk Capacity (AffRC) and the Limit of Reliance. AffRC is 10\% of payroll for 30 years, discounted to a present value using an AffRC discount rate. USS  then sets the Limit of Reliance as 150\% of the AffRC.

\begin{figure}[ht!]

\centering

\includegraphics[width=0.83\textwidth]{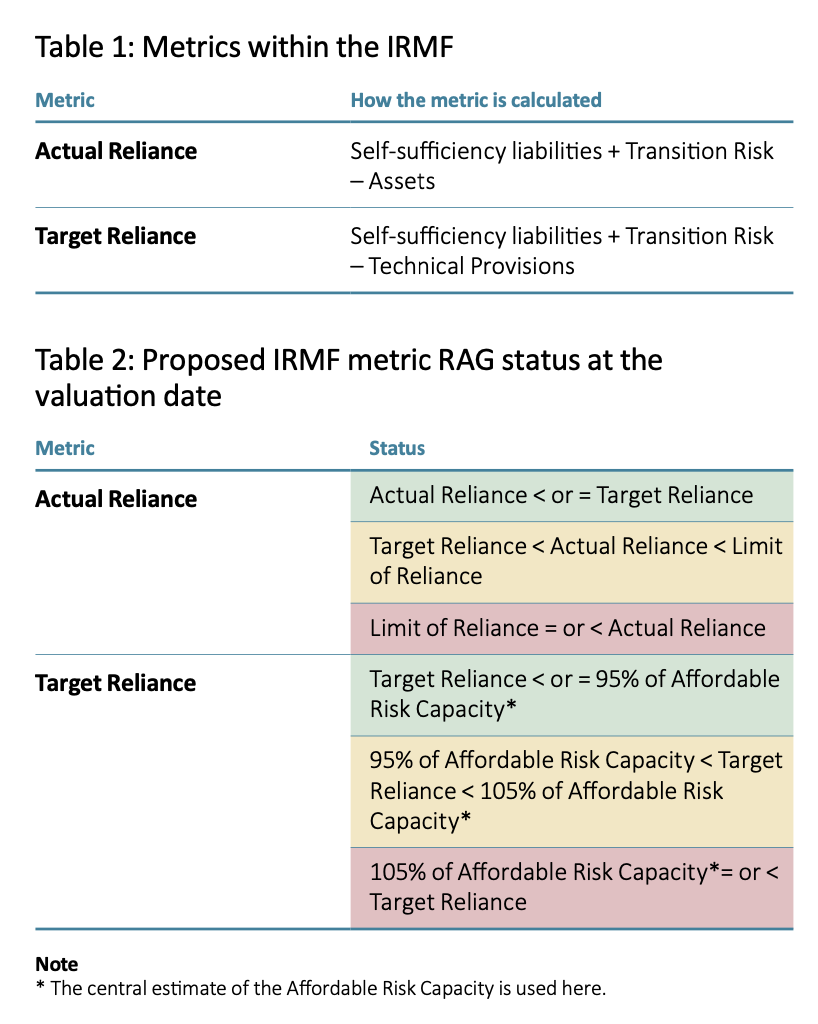}

  \caption{USS metrics Actual and Target Reliance introduced in 2023 \cite{USS_Val_TP_SI_2023}. 
  }
    \label{fig:USS_Tab1_reliance}

\end{figure}

USS has two metrics that aim to measure scheme status in relation to the amount employers could provide via this AffRC. 
These are the Actual and Target Reliance.  USS presented definitions in 2023 as shown in Figure \ref{fig:USS_Tab1_reliance}, which were simplified from the three 2020 metrics A, B and C. All these metrics sit within the USS Integrated Risk Management Framework (IRMF)\footnote{See section \ref{sec:IRMF} for a discussion of the general features of the USS IRMF as well as the relationship between the 2020 metrics A, B and C and the 2023 metrics Actual and Target Reliance. The Glossary \ref{glossary} further covers definitions.}.

\subsection{Actual Reliance can be Green and Red simultaneously}

The Actual Reliance metric is calculated as SfS liabilities minus available scheme funds on a self-sufficiency basis - more precisely, the scheme funds available for the hypothetical self-sufficiency portfolio which are valued as current scheme assets $A$ minus the cost of moving to this portfolio (the Transition Risk $T_\text{Risk}$ which USS cost as \pounds 6-8bn). The liabilities on a SfS basis, the SfS liabilities ($L_\text{SfS}$) are the present value of cashflows due to promised benefits discounted at the SfS discount rate. Hence the Actual Reliance $R_\text{Act}$ is, as per Figure \ref{fig:USS_Tab1_reliance}:
\begin{align}
    R_\text{Act} = L_\text{SfS} - (A - T_\text{Risk}).
\end{align}

The Actual Reliance can be Red, Amber or Green (RAG) as in Figure \ref{fig:USS_Tab1_reliance}. 

First the Green/Amber boundary of the Actual Reliance is considered. The Green status is a condition that Actual Reliance is less than Target Reliance ($R_\text{Tar}$), where Target Reliance is defined with reference to the TP liabilities rather than the Assets, so Target Reliance is written, as per Figure \ref{fig:USS_Tab1_reliance} as: 
\begin{align}
    R_\text{Tar} = L_\text{SfS} - (L_{\text{TP}} - T_\text{Risk}). \label{eq:R_tar} 
\end{align}

The condition for Green status of Actual Reliance can now be simplified as follows: 
\begin{align}
R_\text{Act} &\leq    R_\text{Tar} \nonumber\\
 L_{\text{SfS}}  -( A  -T_{\text{Risk}}) &\leq    L_{\text{SfS}}  -( L_{\text{TP}}  -T_{\text{Risk}})
 \nonumber\\
    A-L_{\text{TP}} & \geq 0, \qquad \text{[ $R_\text{Act}$ Green status]}
\end{align}
where $A-L_{\text{TP}} $ is the Technical Provisions surplus. 

So the Green status for Actual Reliance means the Technical Provisions surplus is greater than or equal to zero. If the TP surplus moves to a TP deficit the Actual Reliance metric turns from Green to Amber. 

Next the Amber/Red boundary is considered, where Limit of Reliance is $R_\text{Lim}$. The condition for Red status of Actual Reliance can be simplified as follows: 
\begin{align}
  R_\text{Act}  &\geq   R_\text{Lim} \nonumber\\
  L_{\text{SfS}} - (A -T_{\text{Risk}} )     &\geq   150\% \text{AffRC} \nonumber\\
    A-L_{\text{SfS}}  & \leq - (150\% \text{AffRC}-T_{\text{Risk}}), \qquad \text{[$R_\text{Act}$ Red status]} \label{eq: Act Rel Red}
\end{align}
where $A-L_{\text{SfS}}$ is the self-sufficiency surplus, or minus the SfS deficit. The Actual Reliance status is then Red if the SfS deficit is greater than the Limit of Reliance on the sector minus the Transition Risk. 

This means the Actual Reliance Green status is a condition that the TP surplus be positive. Independently, the Actual Reliance Red status is  a separate condition on the SfS deficit. So if the scheme records a TP surplus, but a significant SfS deficit the metrics can be simultaneously Green (for TP surplus) and Red (because the SfS deficit is over the limit set by Eq. \ref{eq: Act Rel Red}). This would happen if the SfS liabilities were sufficiently greater than the assets which were greater than the TP liabilities, which would happen if the SfS discount rate were sufficiently below the TP discount rate. The condition for simultaneous Green and Red status is:
\begin{align}
  L_\text{{TP}}  &\leq   A \leq  L_{\text{SfS}}- (150\% \text{AffRC}-T_{\text{Risk}}).
\end{align}

\subsection{Target Reliance sets TP liabilities from SfS liabilities}\label{sec:Tar_rel}

\begin{figure}[hb!]

\centering

\includegraphics[width=0.89\textwidth]{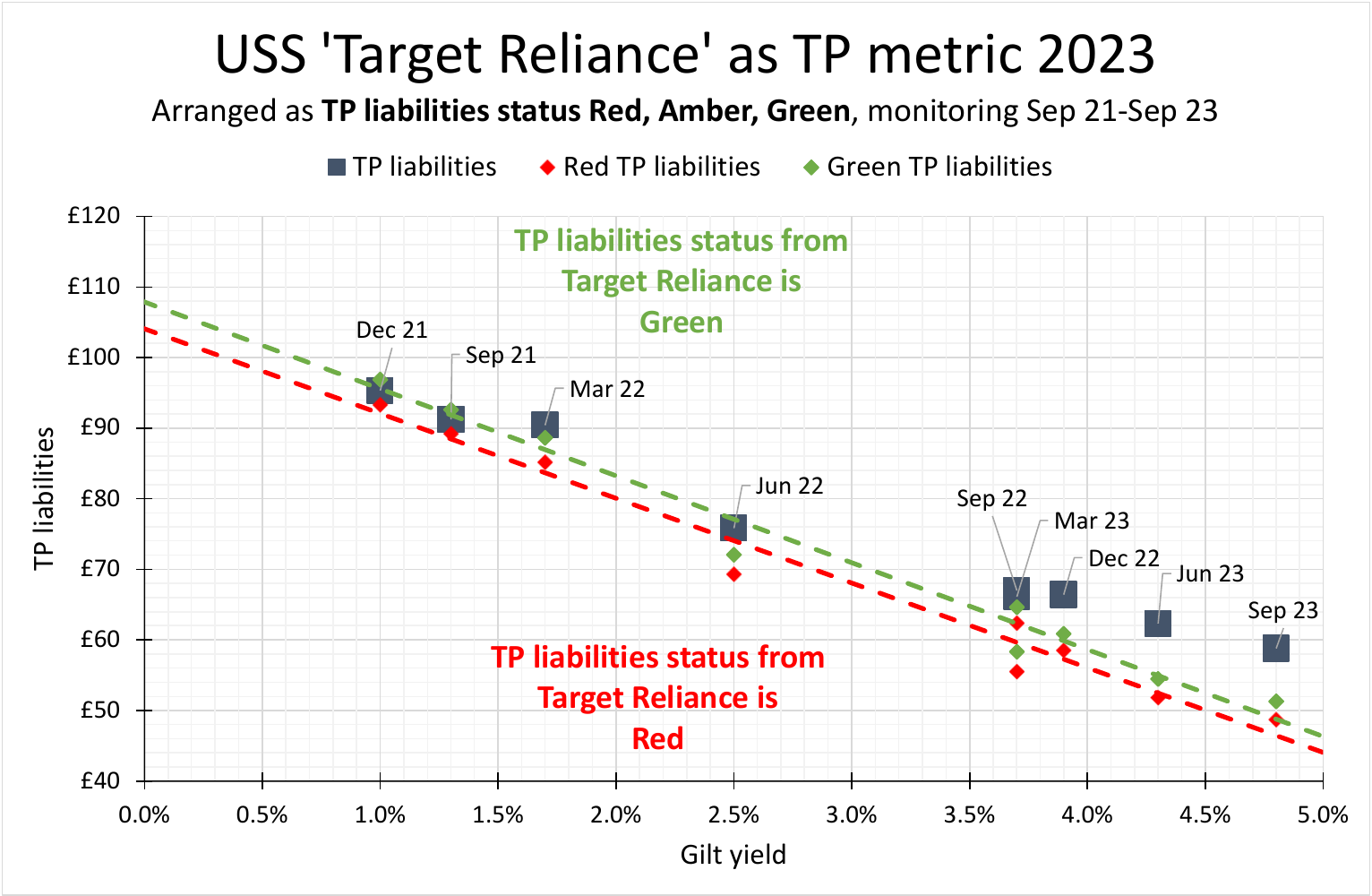}

  \caption{USS Target Reliance  
   as conditions on TP liabilities given by Eqs. \ref{eq:G L_TP} and \ref{eq:R L_TP}. The green dashed line is a linear extrapolation of the Green/Amber boundary. Similarly the red dashed line for the Amber/Red boundary. The TP liabilities (black squares) range from \pounds 60-90bn. If TP liabilities are large the Target Reliance is Green, as the TP liabilities decreases Target Reliance turns Red.
  } 
    \label{fig:TP Tar_Rel_Gilt}

\end{figure}

Target Reliance is defined by USS as `the level of reliance we are aiming to be below, when the scheme is fully funded on the Technical Provisions basis.' Putting aside how it is expected to operate when the scheme is not fully funded, the status, as given in Table \ref{fig:USS_Tab1_reliance} can be Red, Amber or Green (RAG) as follows: 
\begin{align}
R_{\text{Tar}} &\leq  95\% \text{AffRC} \qquad \text{[$R_\text{Tar}$ Green status]},  \label{eq:G Trel}\\
R_{\text{Tar}} &\geq  105\% \text{AffRC}  \qquad \text{[$R_\text{Tar}$ Red status]}. \label{eq:R Trel}
\end{align}
It is notable that the AffRC, which might be expected to be constant, has varied between \pounds 25-35bn over the last two years (since USS monitoring of AffRC started) decreasing as the gilt yield increases. This can be accounted for by the high correlation, $R^2=97\%$, of the AffRC discount rate with the gilt yield \cite{SussexUCUgithub}.

However, the most important point to note is that the 2023 Target Reliance definition of Eq. \ref{eq:R_tar} (and similarly Metric A of 2020) uses SfS liabilities as set by the SfS definition and a Transition Risk which is roughly a constant value. So the RAG status of $R_{\text{Tar}}$ is best conceptualised as measuring the status of the TP liabilities relative to the SfS liabilities plus Transition Risk minus AffRC.

This equivalent Red/Amber/Green status of TP liabilities can be found by substituting Eq. \ref{eq:R_tar} for $R_{\text{Tar}}$ into Eq \ref{eq:G Trel} and \ref{eq:R Trel} for the Green and Red status, and rearranging as conditions on the TP liabilities, to produce the conditions:
\begin{align}
    L_\text{TP}  &\geq  L_\text{SfS}  +T_{\text{Risk}}  - 95\% \text{AffRC} \qquad \text{[$R_\text{Tar}$ Green status]}, \label{eq:G L_TP}\\
    L_\text{TP}  &\leq  L_\text{SfS}  +T_{\text{Risk}}  - 105\% \text{AffRC} \qquad \text{[$R_\text{Tar}$ Red status]}\label{eq:R L_TP}, 
\end{align}
where the inequalities have changed direction as $\text{AffRC}$ now takes a negative sign. 

Figure \ref{fig:TP Tar_Rel_Gilt} shows the $R_{\text{Tar}}$  status applied to TP liabilities. Green/Amber and Amber/Red boundaries as per Eq.\ref{eq:G L_TP} and \ref{eq:R L_TP} are shown as dashed lines. The $R_{\text{Tar}}$ is Green if TP liabilities are large and Red if they are small.  
The actual TP liabilities, set by USS's choice of TP discount rate are shown as black squares, and clearly follow the trend of the metric. So $R_{\text{Tar}}$ acts as a pressure to \textit{increase}  TP liabilities to within a distance of the AffRC from the SfS liabilities plus the Transition Risk. The Target Reliance tethers the TP liabilities directly to the SfS liabilities.

The next section considers how all these aspects link together in the USS Integrated Risk Management Framework to drive the gilt yield dependence.

\newpage 

\section{USS's Integrated Risk Management Framework}\label{sec:IRMF}

\begin{figure}[hb!]

\centering

\includegraphics[width=0.90\textwidth]{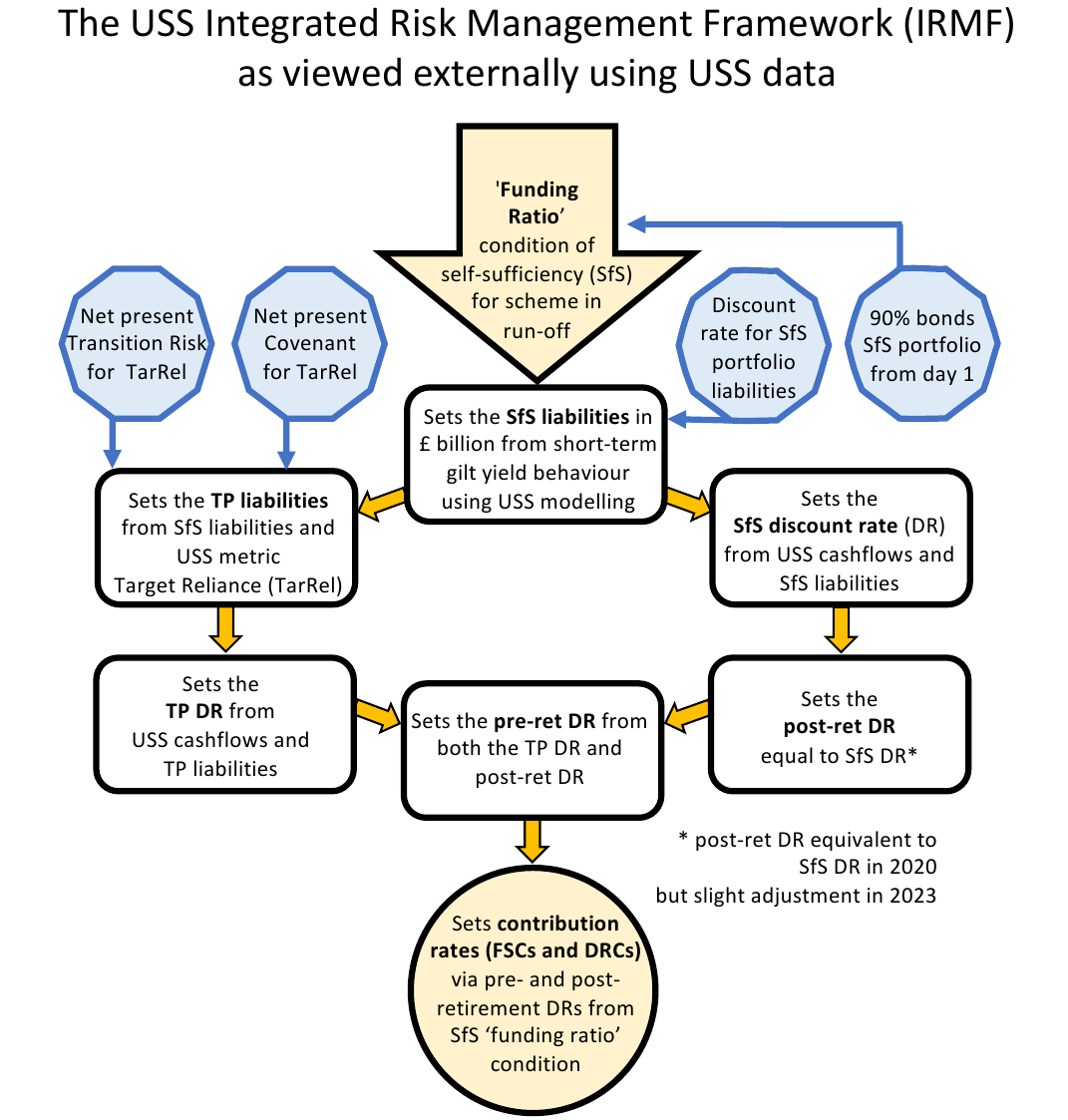}

  \caption{A suggested diagram of the USS Integrated Risk Management Framework from consideration of the analysis in this paper. The funding ratio condition, which depends on the gilt yield, sets the discount rates which set contributions.   
  } 
    \label{fig:IRMF_Diagram_Mechanism}

\end{figure}

USS claims that their valuation approach takes a prudent margin from their best estimates of asset returns, then `iterates' according to their Integrated Risk Management Framework (IRMF). There is no public USS document specifically explaining the USS IRMF, but the analysis in this paper suggests that the IRMF prioritises the funding ratio condition of the USS-specific self-sufficiency definition to inflate costs unnecessarily from the gilt yield, as summarised in Figure \ref{fig:IRMF_Diagram_Mechanism}.

Although there is no single USS IRMF policy, there are two places that contain some useful information. These are the 2023 Technical Provisions Supporting Information, Section 1, \cite{USS_Val_TP_SI_2023} and the 2020 Technical Provisions consultation, Appendix D, \cite{USS_Val_TP_SI_2020}. Both documents discuss the \textit{inputs} into the IRMF including the self-sufficiency definition and, in 2020, the Metrics A, B and C, replaced by Actual and Target Reliance in 2023\footnote{Metric A of 2020 is equivalent to the Target Reliance of 2023. Metrics B and C are equivalent to degrees of the Red/Amber boundary of the Actual Reliance. So the metrics have not substantially changed between 2020 and 2021. Prior to 2018 Test 1 \cite{Marsh_2018} was the primary metric.}. But neither discuss the \textit{mechanism} or iterative process applied to the inputs to achieve satisfactory metrics and discount rates.

USS maintains that it does not target self-sufficiency. Instead they claim that a prudent margin is chosen below the expected return. This is explicit in 
the USS key assumptions for their Actuarial Valuation,  which included the approach of:
\begin{quote}
    ... deriving the discount rates by reference to investment return expectations for the assets the Trustee expects to hold based on the Valuation Investment Strategy... [USS 2023 Actuarial Valuation \cite{USS_vals_mon}]
\end{quote}
However, there is little discussion in USS documents of how the expected returns are considered. The small amount of data on best estimates appear inconsistent. For example, according to USS, the 2020 valuation used a best estimate on the pre-retirement portfolio of either gilts+5.98\%, gilts+5.79\%, gilts+5.28 depending on which USS document is read, see Appendix \ref{app:best-est} . USS does quote the Confidence Level (CL) from the best estimate to the prudently adjusted discount rate. This CL has moved from 65-67\% (2014, 2017 and 2018) to around 80\% (2020) and back to around 70\% (2023). By comparing Figures \ref{fig:pre-ret_unbiased} and \ref{fig:gilt corr pre-ret pru 2020-2023} the CL can be considered an output from the difference between best estimate and prudent returns and the variability  understood from the fact that the best estimate is not highly correlated with the gilt yield, while the pre-retirement discount rate is very  highly correlated.  

Figure \ref{fig:IRMF_Diagram_Mechanism} attempts to fill the absence of a USS explanation of the USS IRMF mechanism. The diagram suggests a plausible mechanism backed by the analysis of this paper. It shows how the self-sufficiency funding ratio dominates the USS-specific SfS definition to set the SfS liabilities as discussed in Section \ref{sec:self sufficiency}. These SfS liabilities then set the post-retirement discount rate. The Target Reliance, which is tethered to the SfS liabilities as discussed in Section \ref{sec:ActTarRel}, then sets the TP DR, and hence also sets the pre-retirement discount rate.  The pre-and post-retirement discount rates then set the contribution rates as discussed in Section \ref{sec:gilt_yield}. In this mechanism the best estimates of returns are bypassed completely. Contribution rates then show the almost perfect correlation with gilt yield (also described in the Section \ref{sec:gilt_yield}) that is ultimately driven by the SfS funding ratio condition.

\newpage
\subsection{USS SfS liabilities and regulatory intervention}\label{subsec:TPR}

Finally the views of the Pensions Regulator (TPR) on the USS valuation and monitoring of self-sufficiency are briefly considered. There are six public letters from TPR to to USS on valuations 2020-2023, dated 26 February, 11 June, 14 July, 24 September 2021, 7 October 2022 and 15 August 2023 \cite{USS_Val_TP_SI_2023,TPR_2021_2022}. 

In June 2021 TPR stated that 
`...in our view, the UUK proposal [to cut benefits while paying 30.7\% contribution rate] is likely to be non-compliant with Part 3 of the Pensions Act 2004.'
They refer to the SfS deficits, TP deficits and covenant strength as being factors they take into consideration. They also stated in July 2021 that they take into account the USS post-valuation monitoring framework: 
\begin{quote}
...continuing with the 2020 valuation was the most measured response to the Scheme’s deteriorating funding position, \textit{in accordance with its monitoring framework... }[emphasis added, TPR to USS, 14 July 2021]
\end{quote}

The August 2023 letter specifically identifies the value of the USS self-sufficiency liabilities as key in TPRs decision regarding both the strength of the covenant and the level of the discount rates:
\begin{quote}
The Scheme’s funding position has improved significantly and that the deficit on a ‘self-sufficiency’ basis has materially reduced since the 2020 valuation (by c.90\%). \textit{The aggregate level of employer resources relative to this much reduced} [SfS] \textit{deficit are characteristic of a Strong covenant.} ... If there was a small surplus or deficit at a future valuation, we would expect the Trustee to consider the \textit{level of TPs, how this compares with the self-sufficiency liabilities, and whether lower discount rate assumptions would be appropriate}. [emphasis added, TPR to USS, 15 August 2023]
\end{quote}

The definition of self-sufficiency used by USS, which includes both the funding ratio condition and the benefit payment condition applied to a 90\% bond portfolio from day one, is specific to USS. Furthermore USS regularly refers to the level of prudence that USS has chosen as being at the limit of acceptability to TPR. Yet it seems clear that the limit set by TPR is taken in the context of high values of the self-sufficiency liabilities that USS themselves have calculated and reported in their monitoring, via the IRMF's use of the funding ratio condition. 

This leads to the question: Is USS inviting intervention from the Pensions Regulator, that argues for high levels of prudence by referencing USS's own calculation of the self-sufficiency liabilities, which are inflated by the funding ratio condition?

\newpage

\section{Summary}\label{sec:summary}

As demonstrated in Section \ref{sec:gilt_yield} the volatility in USS contribution rates over the last decade can almost exclusively (95-99\%) be attributed to changes in the return on UK government bonds (the gilt yield). This has led to excessive prudence and high costs particularly in the years of low gilt yield.
The 2020 valuation occurred at the period of lowest gilt yield, as shown in Figure \ref{fig:BoE gy corr 2000-2024}, and resulted in the highest proposed contribution rates (37\% plus 6.2\% Deficit Recovery Contributions). For the 2023 valuation (carried out when gilt yields had increased following inflation and interest rate increases, and the end of Quantitative Easing \cite{BoE_QE}) contributions for the same benefits had plummeted by more than half, from 43.2\% to 20.6\%.

This change of 22.6ppt in USS proposed contribution rates represents an enormous \pounds 2.3bn a year, to be shared 70:30 between employers and employees, as discussed in Section \ref{sec:summ_gilt}. This astonishing level of volatility in contributions, driven by the very
high dependence of the discount rates on the gilt yield, is not to be expected in a globally diversified pension scheme invested 60\% in equities \cite{Rau_R_historic_returns}.

Importantly the very high correlation of the discount rates with gilt yields is not seen in the USS assumptions for best estimates of returns of the investment portfolios. This would suggest that an underlying mechanism acts within the USS modelling to set the discount rates directly from the gilt yield. It is also possible that a judgement is being made that bypasses the best estimates.  
 
Such an underlying mechanism is proposed in the following four paragraphs. It is supported by analysis in this paper and mapped out in Figure \ref{fig:IRMF_Diagram_Mechanism}

Section \ref{sec:self sufficiency} demonstrated that a little-known or understood `funding ratio condition', contained within the USS self-sufficiency(SfS) definition, is strongly influenced by the gilt yield and produces unnecessarily high self-sufficiency liabilities. Further, this condition does not predict the ability to pay benefits. Preliminary independent analysis,  presented in the same section,  strengthens these conclusions. 
The high level of the SfS liabilities then sets the post-retirement discount rate at a low level.  This is because the post-retirement discount rate is chosen to be equal, or very close, to the SfS discount rate. This leads to the 97-99\% correlation of the post-retirement discount rate with the gilt yield, as seen in Figure \ref{fig:post-ret_pru}. 

Section \ref{sec:ActTarRel} considered the USS metrics of Actual and Target Reliance. Target Reliance is defined as SfS liabilities minus TP liabilities plus the USS estimated cost of moving to SfS. The Target Reliance status of Red, Amber or Green aims to measure whether the cost of moving to self-sufficiency could be borne by employers. The Target Reliance exhibits only a small Amber window, and so 
moves rapidly from the status of Red to Green as the chosen value of the TP liabilities \textit{increases}, as shown in Figure \ref{fig:TP Tar_Rel_Gilt}, setting a \textit{lower bound} on TP liabilities. The choice to place the TP liabilities just above the border of Green Target Reliance tightly couples the TP liabilities to the SfS liabilities and therefore to the gilt yield. The SfS liabilities are inflated by the funding ratio condition, and hence so are the TP liabilities. 
This in turn depresses the TP discount rate which sets the pre-retirement discount rate. The pre-retirement discount rate (from a portfolio of  90\% equities) then demonstrates the astonishing 97-99\% correlation with the gilt yield, seen in Figure \ref{fig:gilt corr pre-ret pru 2020-2023}. A correlation not seen in best estimate returns, Figure \ref{fig:pre-ret_unbiased}.

These pre- and post-retirement discount rates then set contributions for both the Future Service Costs and any deficits. These contribution rates exhibit the same very high correlation with the gilt yield, as seen in Figures \ref{fig:gilt corr val 14-23}, \ref{fig:gilt corr mon  2020 cuts} and \ref{fig:gilt corr mon 2020}. A reminder that this instability has caused enormous changes in proposed contribution rates, of over 20\% of salaries or \pounds 2bn per year, as shown in Table \ref{tab:contributions_breakdown}. It is unsurprising that UK universities have seen a decade of industrial disputes.

The overarching mechanism responsible for USS valuations is the USS Integrated Risk Management Framework (IRMF). 
There is no single policy document on how USS uses their IRMF for valuations, but Figure \ref{fig:IRMF_Diagram_Mechanism} of Section \ref{sec:IRMF} suggests an underlying mechanism. This diagram aims to represent the four paragraphs above, which tie together the data analysis and exploration of self-sufficiency modelling, gilt yield dependence and metrics of Sections \ref{sec:gilt_yield}, \ref{sec:self sufficiency} and \ref{sec:ActTarRel}.  

A further four points are noted: 
(i) The funding ratio condition is applied to a self-sufficiency portfolio of 90\% bonds from day one, even though USS states such a portfolio would take a decade to implement \cite{USS_Transition_Risk_2022} and portfolios with higher allocation of bonds appear more likely to fail to pay pensions. 
(ii) As discussed in Section \ref{sec:IRMF}, USS claims that their valuations use a prudent adjustment from their best estimate of expected returns on their actual investment strategy. Yet the only useful data on these best estimates is the monitoring of Figure \ref{fig:pre-ret_unbiased}.
The little valuation data on the USS choice of best estimates 
is patchy and inconsistent, see Appendix \ref{app:best-est}.
(iii) The Pensions Regulator's interventions, arguing for higher levels of prudence, have specifically referenced the high SfS and TP liabilities, as reported by USS, see Section \ref{subsec:TPR}, suggesting that these invite regulatory concern.
(iv) 
USS public definitions of SfS made no, or minimal, reference to the funding ratio condition until 2023. Stakeholders that did reference SfS (including JEP, UUK and UCL) did not mention the funding ratio condition, see Appendix \ref{app:USS sfs 2023}. 

Finally, as noted above, there is no unifying policy document on USS's Integrated Risk Management Framework. There is no policy on how USS uses expert judgement to interpret modelling to make informed real-world decisions \cite{Erica_Thompson}. There is no policy on producing analysis to a particular standard or a level of access. 

\section{Conclusions}\label{sec:concl}

From the consideration of ten years of public USS data, it 
appears that the USS valuation methodology has produced 
contribution rates with high volatility that can be attributed (95-99\%) directly to movements in the gilt yield. This is despite the scheme being majority invested in growth assets.
In an era of low gilt yields this has led to attempts to close the scheme, huge benefit cuts and a decade of industrial action to successfully save the scheme and restore benefits. As discussed in the Summary \ref{sec:summary}, a source of this gilt yield dependent volatility can be attributed to the USS-specific modelling of self-sufficiency that included a mysterious, but crucially important, funding ratio condition within an overly complex valuation approach. This seemed to set contributions directly from gilt yields through the modelling; or via decision making; or a combination of both.

Can USS move beyond cycles of controversial methodologies, excessive costs and industrial disputes? USS has seen recent changes in key personnel. Their work to accelerate the 2023 valuation facilitated timely restoration of the benefits cut in April 2022, subsequent to agreements between UUK and UCU \cite{UUK_UCU_JS_Oct_2023}. Benefits were fully restored on 1 April 2024 to over 200,000 scheme members along with £900 million of scheme funds allocated to the recovery of benefits lost since 2022. 
The UK Pensions Regulator’s draft DB funding regulations have recently been published following public consultation \cite{TPR_DB_new_code}. The USS response was commendable in its approach to seek, promote and publish stakeholders’ views recognising the open and long term nature of the scheme \cite{USS_Briefings_Analysis}.

A core aim of the analysis in this paper has been to support evidence-based, transparent decision making and consensus building for future valuations. To this end the ongoing work between USS, UUK and UCU to explore stability is also to be welcomed. As a UCU negotiator, on the USS Stability Working Group, I am part of that group and the analysis documented here has formed a major component of the material I have brought to these ongoing discussions.

The next USS valuation is due to run from March 2026. It is, of course, not possible to say yet what approach will be used, although some grounds for optimism may be emerging. As details take shape it will be necessary to engage in all ongoing work to scrutinise methodologies as, and when, they evolve.

There is also an important piece of work needed to investigate more broadly the apparent lack of regulatory oversight of valuation methodologies.
While USS members succeeded in keeping USS open, most UK DB schemes have closed. The pensions landscape, and those schemes' members, are much poorer as a result.

\vspace{0.1cm}
{
\noindent
\hrulefill\hspace{0.2cm} 
\decofourleft\decofourright
\hspace{0.2cm} \hrulefill}

\newpage

\noindent \textbf{Note added post-analysis}

Since completion of the analysis in this pre-print the USS December 2023 monitoring has been published \cite{USS_vals_mon}. The costs of post-April 2022 benefits against gilt yield, Figure \ref{fig:gilt corr mon  2020 cuts}, can be updated to include the December 2023 data and shows a change in correlation from 99.1\% to 99.0\%. The costs of pre-April 2022 (or restored) benefits against gilt yield, Figure \ref{fig:gilt corr mon 2020}, can be updated to include December 2023 data and shows a change in correlation from 95.5\% to 95.8\%. Neither of these updates change any results or conclusions. Both plots are available at github \cite{SussexUCUgithub}.

It is not possible to update other analysis as USS monitoring has changed following the March 2023 valuation, and so a like-for-like comparison is not available for any data except Future Service Costs. For example the discount rates have been reset to start from March 2023 valuation values (which are slightly different to the monitoring values), and best estimates are now presented with respect to Index Linked Gilts (ILG+) although values for ILG+ are not currently available.

\noindent \textbf{Affiliations and acknowledgements} 

I am an elected University College Union (UCU) national negotiator and member of the USS Joint Negotiating Committee. I am the University of Sussex and Institute of Development Studies (IDS) UCU Pensions and Finance Officer. 

Thanks and appreciation for helpful discussions and comments on drafts of this manuscript to Dooley Harte, Mark Hindmarsh, Sarah Joss, Con Keating, Sam Marsh, Mike Otsuka, Mark Taylor-Batty, Woon Wong. Thanks to colleagues past and present on the UCU Superannuation Working Group and the Sussex and IDS Pensions Working Group. Thanks to useful conversations with First Actuarial's Sarah Abraham and Derek Benstead and colleagues representing UUK, USS and AON, in particular Anthony Odgers and John Coulthard. All errors my own.

\newpage

\bibliographystyle{unsrt}
\bibliography{USS_bib.bib}

\newpage

\appendix
\section{Appendices}

These appendices cover gilt yield dependence \ref{app:gilt yield}, USS and other definitions of self-sufficiency 2014-2023 \ref{app: self-sufficiency definitions}, inconsistencies in USS data on best estimates \ref{app:best-est} and a glossary \ref{glossary}.
All data at 
\href{https://github.com/SussexUCU/USS_vals_rev}{https://github.com/SussexUCU/USS\_vals\_rev}.

\subsection{Gilt yield dependence 2014-2023}\label{app:gilt yield}

\subsection*{Future Service Cost gilt yield dependence}

As covered in Section \ref{sec:gilt_yield} Future Service Costs (FSCs) can be determined to a very high degree ($>$95\%) from the gilt yield for all valuations over the last decade and for all monitoring since March 2020. All gilt yield values used in the analysis are either from USS or the publicly available data from Bank of England (BoE) \cite{BoE-database}. 

\begin{figure}[hb!]

\centering

\includegraphics[width=0.9\textwidth]{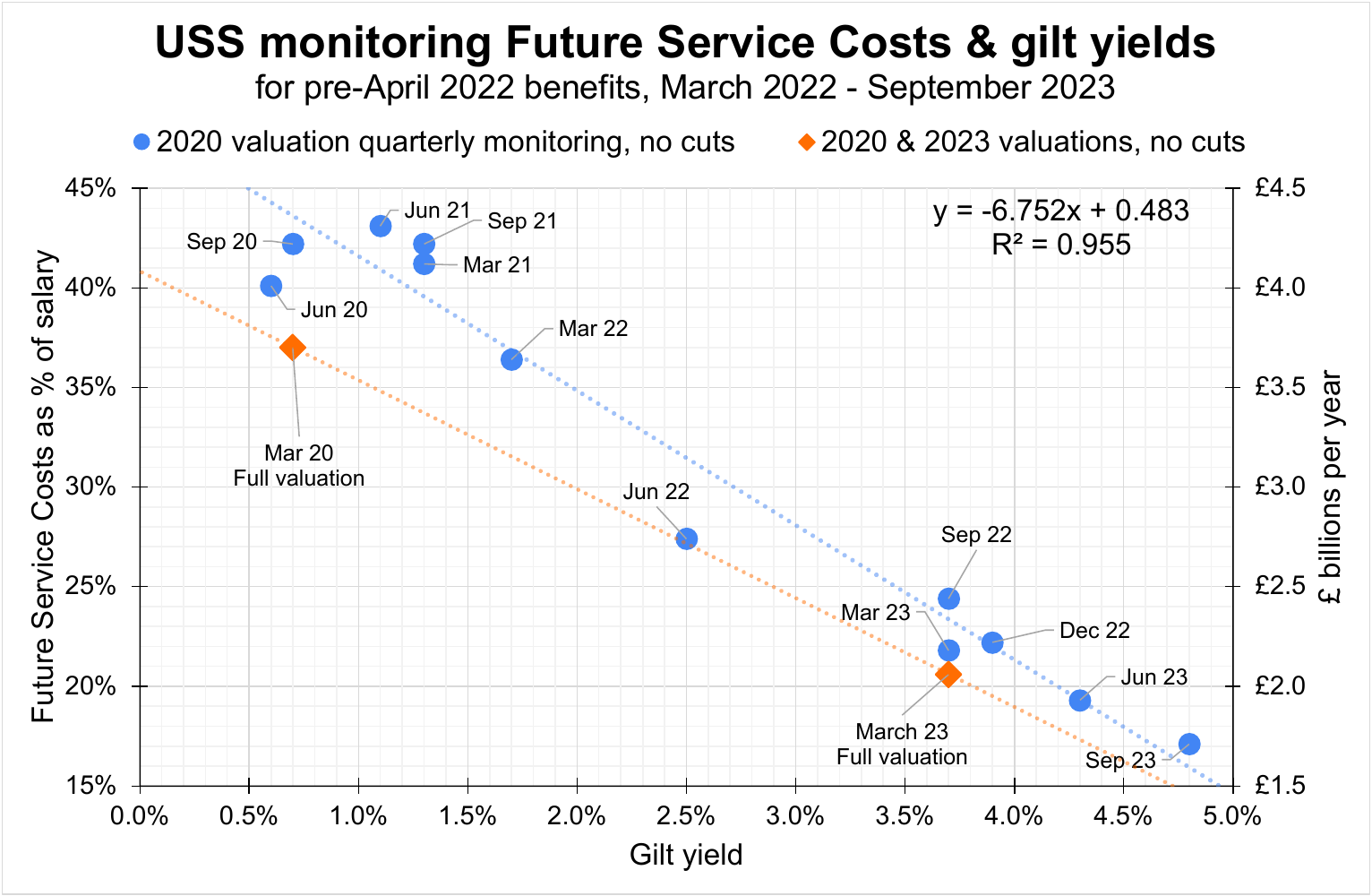}

  \caption{USS quarterly monitoring, March 2020 to September 2023, shows a 95.5\% correlation between FSCs and gilt yield. The FSCs are shown for pre-2022 benefits (no cuts). The 2020 and 2023 valuation FSC are also shown. Valuations and monitoring for post-2022 benefits (with cuts) can be seen in Figures \ref{fig:gilt corr val 14-23} and \ref{fig:gilt corr mon  2020 cuts}. }

    \label{fig:gilt corr mon 2020}

\end{figure}

The three figures Figures \ref{fig:gilt corr val 14-23}, \ref{fig:gilt corr mon  2020 cuts} and \ref{fig:gilt corr mon 2020} all show the remarkably high correlation between FSCs and gilt yields for all five valuations in the last decade, and then the monitoring from 2020 for both cut (post-2022) and uncut (pre-2022) benefits.  

Table \ref{tab:val14-23 mon 2020} shows the correlations and linear equations for data as presented in Figures \ref{fig:gilt corr val 14-23}, \ref{fig:gilt corr mon  2020 cuts} and \ref{fig:gilt corr mon 2020}. Correlations are all between 95-99\% with the linear equations demonstrating that post-2020 benefit structure produces lower FSCs with lower dependence (gradient) on the gilt yield. Monitoring is more prudent (higher intercept of 48\% and steeper gradient of -6.8) than the valuations of 2014-2023 (with intercept 40\% and gradient -5.4).

\begin{table}[h]
    \centering
\begin{tabular}{|l|l|l|}

\hline
 \textbf{Types of USS data }& \textbf{R-squared} &  \textbf{Linear equation}   \\
\hline
\hline
 All valuations 2014 - 2023  & 97.6\% &  $y=-5.4x+40\%$  \\
(pre-2022 benefits) & &    \\
\hline
 Monitoring Mar 2020 - Sep 2023 & 99.1\% &  $y=-3.5x+31\%$   \\
 (post-2022 benefits) & &    \\
\hline
Monitoring Mar 2020 - Sep 2023 & 95.5\% & $y=-6.8x+48\%$   \\
(pre-2022 benefits) & &    \\
\hline

\end{tabular}

\caption{Correlations and linear equations for FSCs dependence on gilt yields from USS valuations 2014-2023 and USS post-2020 monitoring for pre-2022 (without cuts) and post-2020 (with cuts) benefit structures.}
    \label{tab:val14-23 mon 2020}
\end{table}

USS monitoring following the valuations of 2014, 2018 and 2020 valuations is next considered, using lower quality data. This monitoring is shown to have the same features and similar correlations shown in Table \ref{tab:val14-23 mon 2020}. 

\subsubsection*{USS monitoring of FSCs 2014-2022} 

\begin{figure}[h!]

\centering

\includegraphics[width=0.9\textwidth]{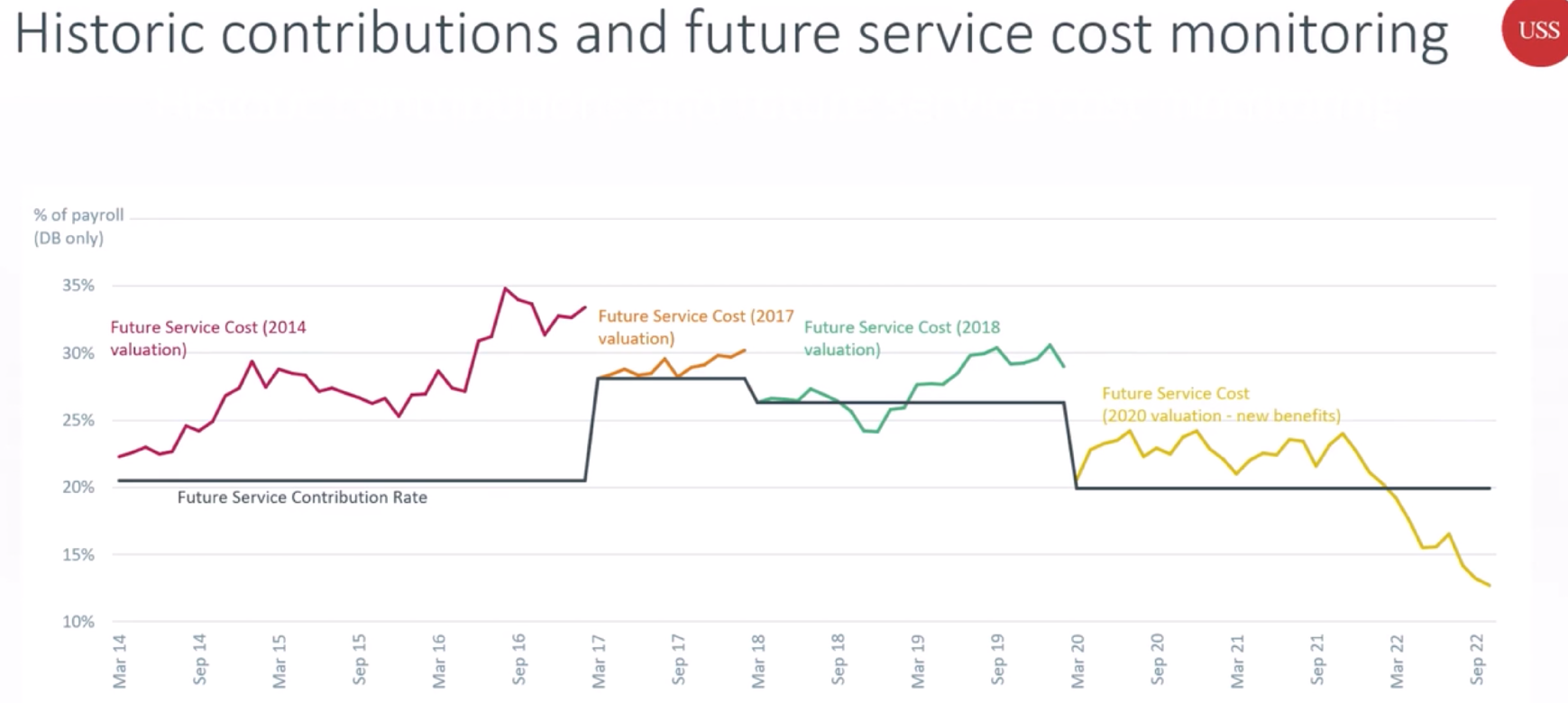}

  \caption{Image of data presented at USS Institutions Meeting 2022 by USS CEO at the time \cite{USS_ins_meet_2022}. The FSC monitoring is for the DB scheme only. 
  In the absence of other published sources, a data grab \cite{Rohatgi2022} was used to obtain monthly USS monitoring of FSC values from 2014 to 2022.
  }
    \label{fig:BillGraph2022}

\end{figure}

\begin{figure}[hb!]

\centering

\includegraphics[width=0.9\textwidth]{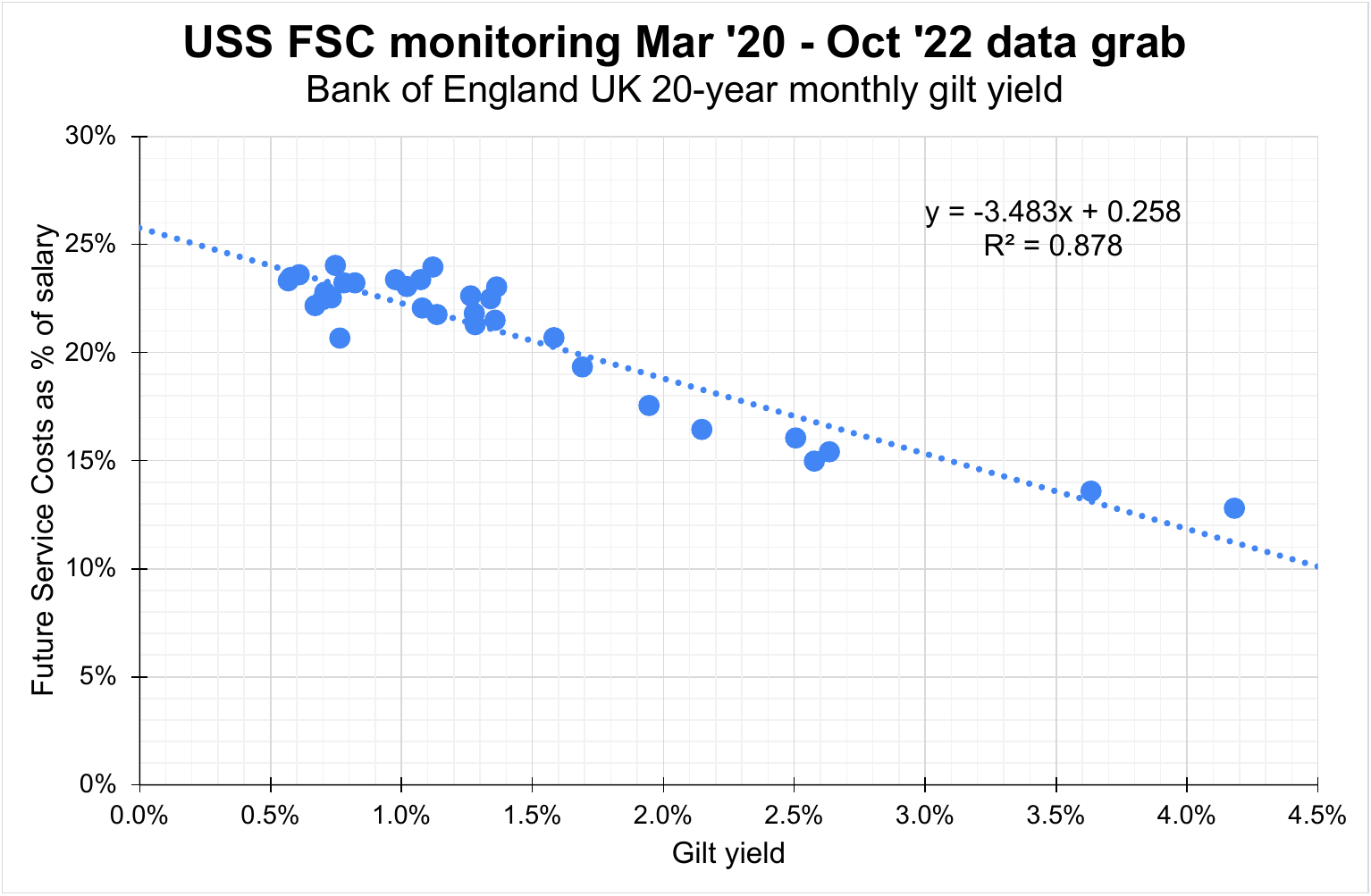}

  \caption{USS monitoring data of Fig. \ref{fig:BillGraph2022}. FSCs for DB only from March 2020 to October 2022 for post-2022 benefits structure (with cuts) and BoE 20-year gilt yields. An 88\% correlation is found between FSC and gilt yields.  
  }
    \label{fig:2020_2022_bill}

\end{figure}

\begin{table}[h]
    \centering
\begin{tabular}{|l|l|l|}

\hline
 \textbf{Types of USS data }& \textbf{R-squared} &  \textbf{Linear equation}   \\
\hline
\hline
Monitoring Mar 2014 - Feb 2018  & 89.6\% &  $y=-6.2x+43\%$  \\
(pre-2022 benefits) & &    \\
\hline
Monitoring Mar 2018 - Feb 2020 & 85.6\% &  $y=-4.7x+35\%$   \\
(pre-2022 benefits) & &    \\
\hline
Monitoring Mar 2020 - Oct 2022  & 87.8\% & $y=-3.5x+26\%$   \\
(post-2022 benefits) & &    \\
\hline

\end{tabular}

\caption{Correlations and linear equations for monthly USS FSCs monitoring following the three implemented valuations of March 2014, March 2018 and March 2020, data from  Figure \ref{fig:BillGraph2022} and Bank of England 20-year gilt yields.}
    \label{tab:2014-2022_mon}
\end{table}

There are no publicly available monitoring data for the period 2014-2020, but a graph, shown below in Figure \ref{fig:BillGraph2022}, was presented at the USS Institutions Meeting, London, UK in 2022. The very high volatility of FSC monitoring data (13-35\%) is obvious from Figure \ref{fig:BillGraph2022}. The results of correlating the FSCs from a data grab \cite{Rohatgi2022} of the image with Bank of England gilt yields are presented in Table \ref{tab:2014-2022_mon}. 

A direct comparison with the data of Figures \ref{fig:gilt corr val 14-23}, with the data in Figure \ref{fig:BillGraph2022} is challenging for several reasons. Figure \ref{fig:BillGraph2022} shows FSCs for DB scheme only, unlike the data in Figures \ref{fig:gilt corr val 14-23}, \ref{fig:gilt corr mon  2020 cuts} and \ref{fig:gilt corr mon 2020} which show FSCs for DB + DC components.  

A second challenge is the low quality of data in Figure \ref{fig:BillGraph2022}, but it is possible to data grab from the image. A third is the lack of USS reported gilt yield values, but Bank of England (BoE) 20-year gilt yields can be used and consistency checked with known data.

The results of the data grab of FSCs from Figure \ref{fig:BillGraph2022} correlated with BoE gilt yield Figure \ref{fig:BoE gy corr 2000-2024} is shown in Figure \ref{tab:2014-2022_mon} for the monitoring period 2014-2017.

The monitoring from 2020 onward (post-2022 benefits) is the only range contained in both the high quality data of Figures \ref{fig:gilt corr val 14-23}, \ref{fig:gilt corr mon  2020 cuts} and \ref{fig:gilt corr mon 2020} and the lower quality data of Figure \ref{fig:BillGraph2022}. So it will be useful to compare the key features from these two different data sets, and this is shown in Table \ref{tab:compBill_USSmoncorr}.

\begin{table}[h]
    \centering
\begin{tabular}{|l|l|l|}

\hline
 \textbf{Monitoring 2020-2023}& \textbf{R-squared} &  \textbf{Linear equation}   \\
  \textbf{(post-2022 benefits)}&  &    \\
\hline
\hline
FSCs and gilt yields from Fig. \ref{fig:gilt corr mon 2020}   & 99.1\% &  $y=-3.5x+31\%$  \\
DB+DC scheme & &    \\
\hline
FSCs and gilt yields from Fig. \ref{fig:2020_2022_bill} and BoE & 87.8\% &  $y=-3.5x+26\%$   \\
DB only & &    \\
\hline

\end{tabular}

\caption{Comparison of Figure \ref{fig:gilt corr mon  2020} data with Figure \ref{fig:2020_2022_bill} data for the same range of USS monitoring 2020-2023 (post-2022 benefits).}
    \label{tab:compBill_USSmoncorr}
\end{table}

Comparing the linear equations in Table \ref{tab:compBill_USSmoncorr}, the gradients (-3.5) give good agreement. The y-axis intercept (corresponding to hypothetical FSC contribution at 0\% gilt yield) is 5\% higher for the DB+DC scheme data (31\%) than the DB only data (26\%). This corresponds to a uniform upwards shift of 5\% from the DB only to the DB+DC data, and is in good agreement with the 5.3\% DC contribution for post-2022 benefits, as reported by USS in the 2020 Actuarial Valuation \cite{USS_vals_mon}. 

The R-squared value is 99.1\% for the higher quality USS only data and 87.8\% for the lower quality data. This lower correlation for the lower quality data, from using data grab software \cite{Rohatgi2022} from an image of Figure \ref{fig:BillGraph2022}, is to be expected. 

But in general the same features emerge between the high quality data of USS valuations and post-2020 monitoring compared with the lower quality post-2014 monitoring to a consistency to two significant figures.

\newpage 

\subsection*{Technical Provisions gilt yield dependence}

\begin{figure}[hb!]

\centering

\includegraphics[width=0.9\textwidth]{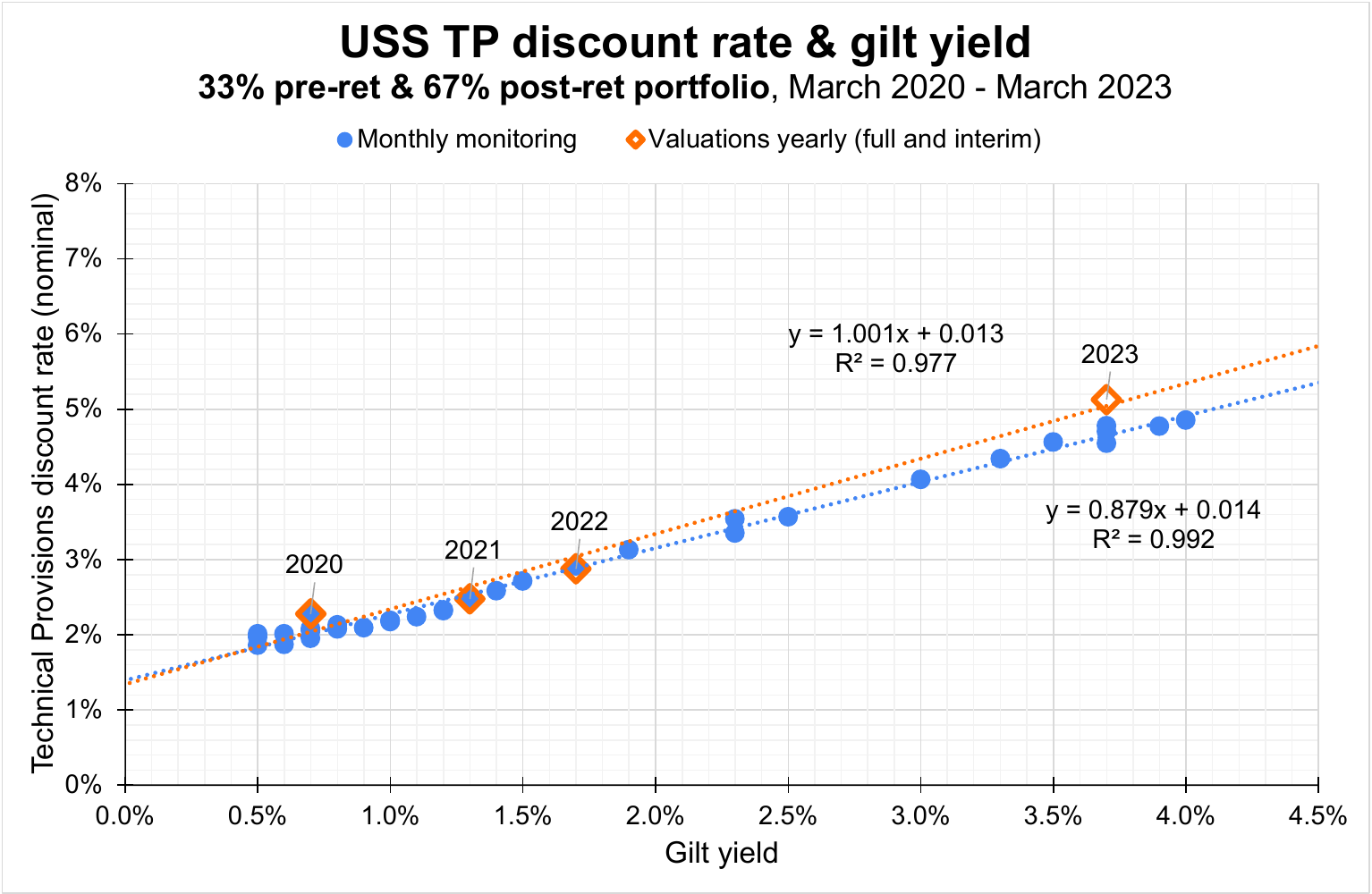}

  \caption{The Technical Provisions discount rate is 33\% pre-retirement portfolio and 67\% post-retirement portfolio. By combining these portfolios a 98\% and 99\% correlated of the TP discount rate with the gilt yield is found for monthly monitoring and full and interim valuations between March 2020 and March 2023.
  }
    \label{fig:TP DR}

\end{figure}

This section considers the Technical Provisions (TP) liabilities and discount rates. This is because USS does not produce regular monitoring data for Deficit Recovery Contributions (DRCs) so it is not possible to analyse DRCs directly. There are however two sources of data that are closely associated with the DRCs. These are pre- and post-retirement discount rates (pre- and post-ret DR) that make up the TP discount rate and hence determine the TP liabilities\footnote{ Section \ref{subsec:TP DR gilt yield} considers how DRCs related to pre- and post-ret DR and TP liabilities.}. Figure \ref{fig:TP DR} shows the TP discount rate calculated from Figures \ref{fig:gilt corr pre-ret pru 2020-2023} and \ref{fig:post-ret_pru}.

\begin{figure}[ht!]

\centering

\includegraphics[width=0.9\textwidth]{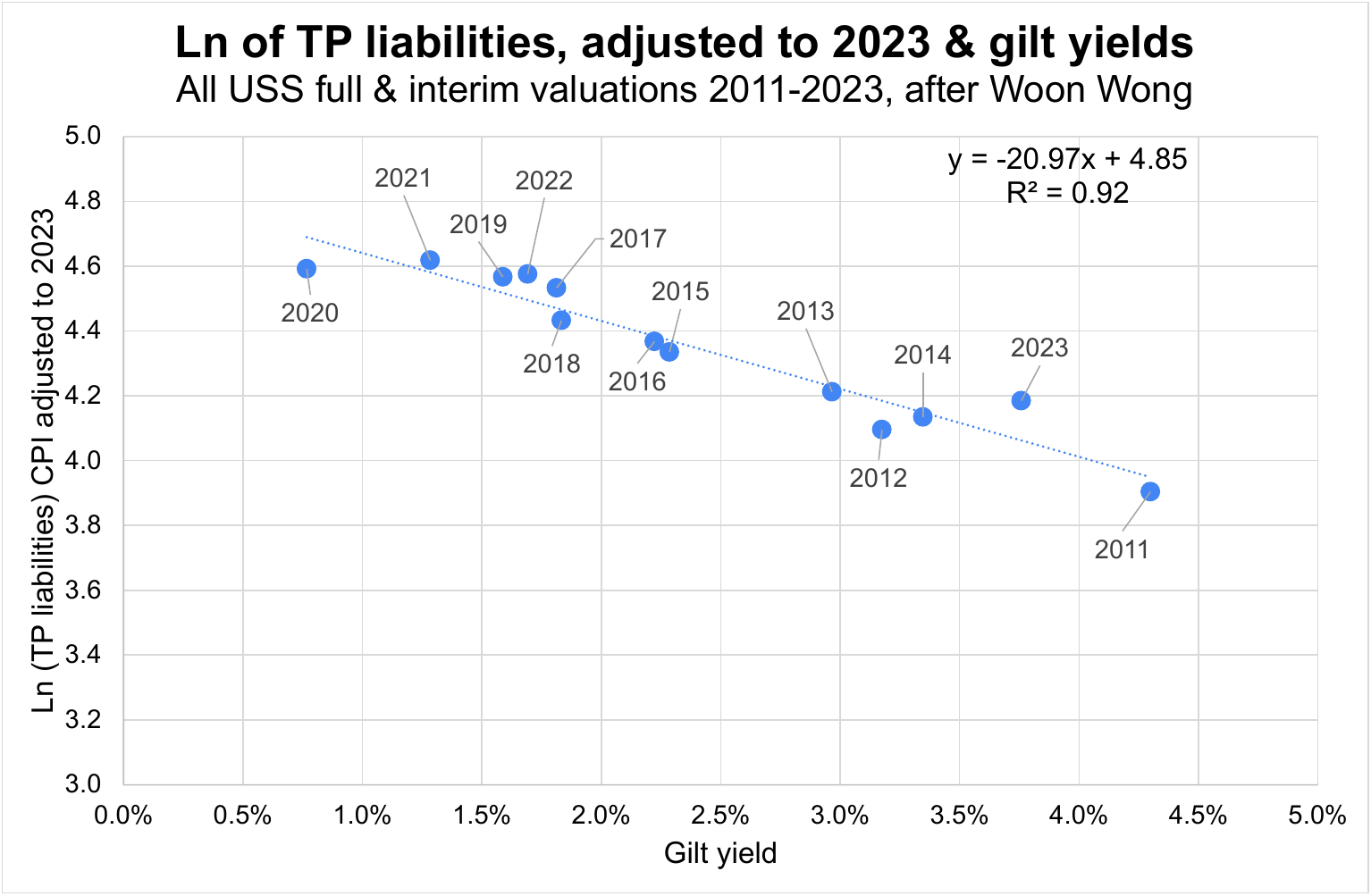}

  \caption{Following Wong \cite{Woon_CDF_2018/12} the Natural logarithm of Technical Provisions liabilities against gilt yield 2011-2023. The correlation is $92\%$. 
  }
    \label{fig:Ln TP BoE}

\end{figure}

The Technical Provisions liabilities are considered next. In 2018 Wong  demonstrated a 99\% correlation of the natural logarithm ($\ln{}$) of the Technical Provisions liabilities against Index Linked 20-year gilt yield (IL20) \cite{Woon_CDF_2018/12}. All full and interim valuations 2011-2018 were included. The natural logarithm was taken as a linear expansion in terms of the discount rate. Using the  Bank of England (BoE) 20-year gilt yields instead of IL20 produces a similarly high correlation of 97\% for the same range of valuations. 

Updating this work to include all full and interim valuations 2011-2023, and (as per Wong) adjusting the value of the TP liabilities to 2023 values using ONS CPI to 2023 values for the TP liabilities produces Figure \ref{fig:Ln TP BoE}. A correlation of 92\% is produced when plotting the ln of TP liabilities against the BoE gilt yield. Using IL20 instead of BoE gives a 95\% correlation. However, BoE has been used through out this paper as this data is readily publicly available.

\subsection*{Bank of England 20-year conventional gilt yields}

USS publish quarterly and sometimes monthly monitoring reports alongside yearly full or interim valuations \cite{USS_vals_mon}. The data included in these reports varies but since March 2020 has included the USS choice of single equivalent (SE) gilt yield. 

\begin{figure}[hb!]
\centering
\includegraphics[width=0.86\textwidth]{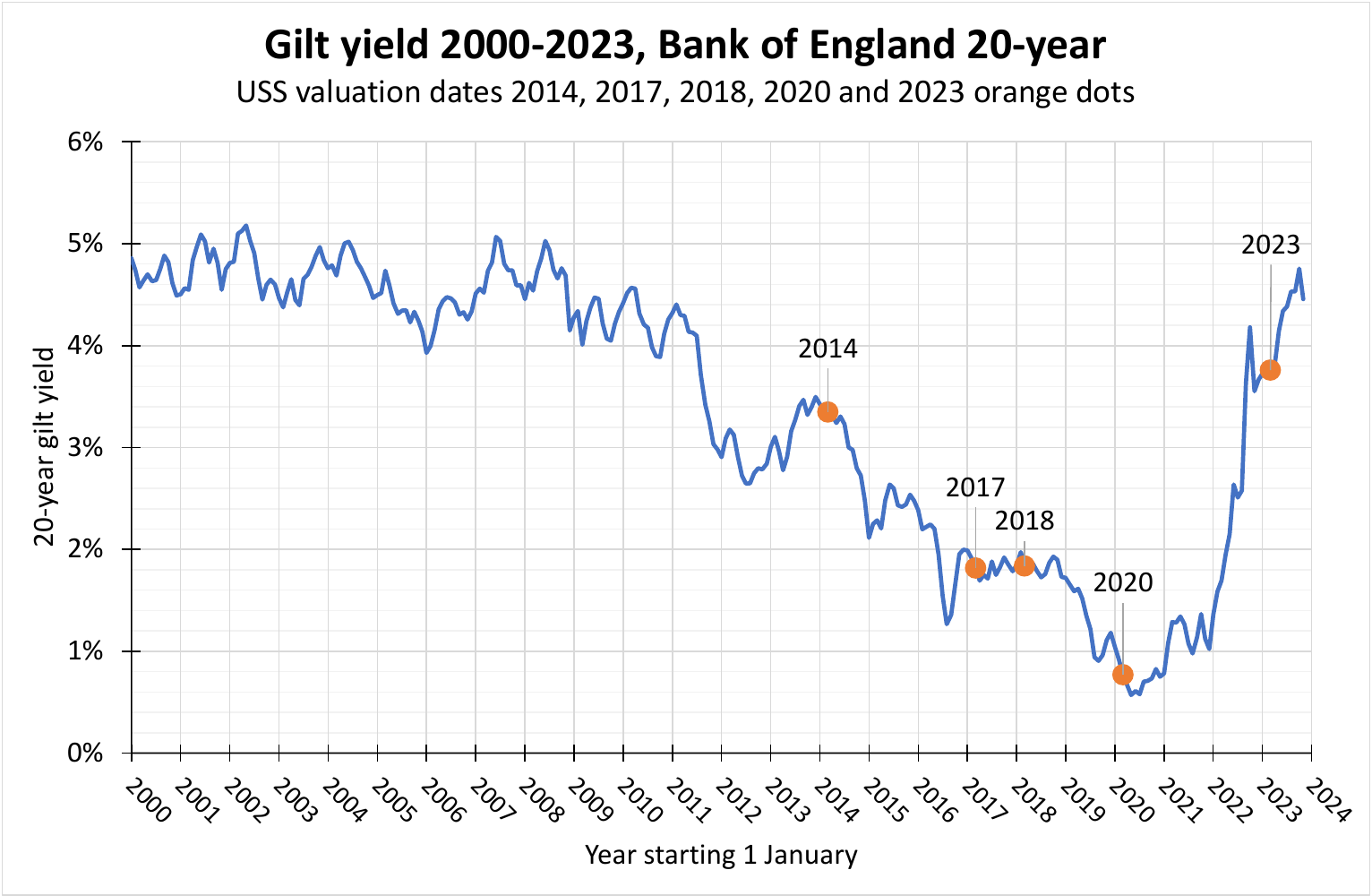}

  \caption{Bank of England 20-year gilt yield 2000-2023, data set IUMALNPY \cite{BoE-database} with a decade of USS valuations dates 2014-2023 shown as orange dots.
  }
    \label{fig:BoE gy corr 2000-2024}

\end{figure}

However, USS SE gilt yield has not been published for all full and interim valuations. Where such data are not available the Bank of England (BoE) 20-year gilt yield values are used instead, data set IUMALNPY from \cite{BoE-database}, and these are publicly available and free. Table \ref{tab:USSBoEComp} compares the gilt yield values chosen by USS with this BoE gilt yield for quarterly data March 2020 to March 2023. The agreement is very good and mean difference is 0.009\% with a standard deviation of 0.1\%. Figure \ref{fig:BoE gy corr 2000-2024} shows BoE gilt yields since 2020 with valuation dates 2014-2023.
\begin{table}[!ht]
    \centering
    \begin{tabular}{|r|r|r|r|r|}
    \hline
        Date & USS gilt yield & BoE gilt yield & Difference \\ \hline \hline
        Mar 2020 & 0.7\% & 0.766\% & 0.07\% \\ \hline
        Jun 2020 & 0.6\% & 0.610\% & 0.01\% \\ \hline
        Sep 2020 & 0.7\% & 0.709\% & 0.01\% \\ \hline
        Dec 2020 & 0.6\% & 0.748\% & 0.15\% \\ \hline
        Mar 2021 & 1.3\% & 1.283\% & -0.02\% \\ \hline
        Jun 2021 & 1.1\% & 1.265\% & 0.17\% \\ \hline
        Sep 2021 & 1.3\% & 1.136\% & -0.16\% \\ \hline
        Dec 2021 & 1.0\% & 1.021\% & 0.02\% \\ \hline
        Mar 2022 & 1.7\% & 1.692\% & -0.01\% \\ \hline
        Jun 2022 & 2.5\% & 2.634\% & 0.13\% \\ \hline
        Sep 2022 & 3.7\% & 3.634\% & -0.07\% \\ \hline
        Dec 2022 & 3.9\% & 3.666\% & -0.23\% \\ \hline
        Mar 2023 & 3.7\% & 3.758\% & 0.06\% \\ \hline
    \end{tabular}

      \caption{USS and Bank of England quarterly single equivalent 20-year gilt yield values March 2020 to March 2023, data set IUMALNPY \cite{BoE-database}. The difference is also shown. The mean difference is 0.009\% with a standard deviation of 0.1\%.
  }
\label{tab:USSBoEComp}\end{table}

\newpage

\subsection{USS and other self-sufficiency definitions 2014-2023}\label{app: self-sufficiency definitions}

This appendix documents definitions of self-sufficiency. See also Section \ref{sec:self sufficiency}. 

\textbf{2023 USS SfS definition}: 
has the following detail on the funding ratio 
\cite{USS_Val_TP_SI_2023}.

\begin{quote}

\small\textbf{1.2 Self-sufficiency as our benchmark for risk}\label{app:USS sfs 2023}

If there were no covenant, to provide benefit security, the trustee has previously concluded that the scheme would need to be funded to at least a self-sufficiency level. Self-sufficiency (as defined by the trustee) is a low-risk investment strategy for funding the scheme in the absence of a covenant.

It corresponds to a confidence level of 95\% (equivalent to a 5\% failure rate) of passing the following tests without the need for any additional contributions:

\begin{enumerate}
    \item Being able to pay all benefits when they fall due(that is,not exhausting all capital before the final benefit is paid).
    \item Not falling below a 90\% self-sufficiency funding level at each triennial valuation.
\end{enumerate}

The resulting notional investment strategy is constructed to meet these criteria via the following principles:

\begin{itemize}
\item The strategy should be well-hedged (above 90\%) against interest rate and inflation risk, and it should retain sufficient collateral to support any leverage.
\item The strategy must be able to organically meet cash flows, allowing for
periodic rebalancing.
\item  The strategy must provide a reasonable return margin over gilts.
\end{itemize}

For the 2023 valuation, the trustee has adopted the following notional self-sufficiency investment strategy:

\begin{figure}[hb!]

\centering

\includegraphics[width=0.85\textwidth]{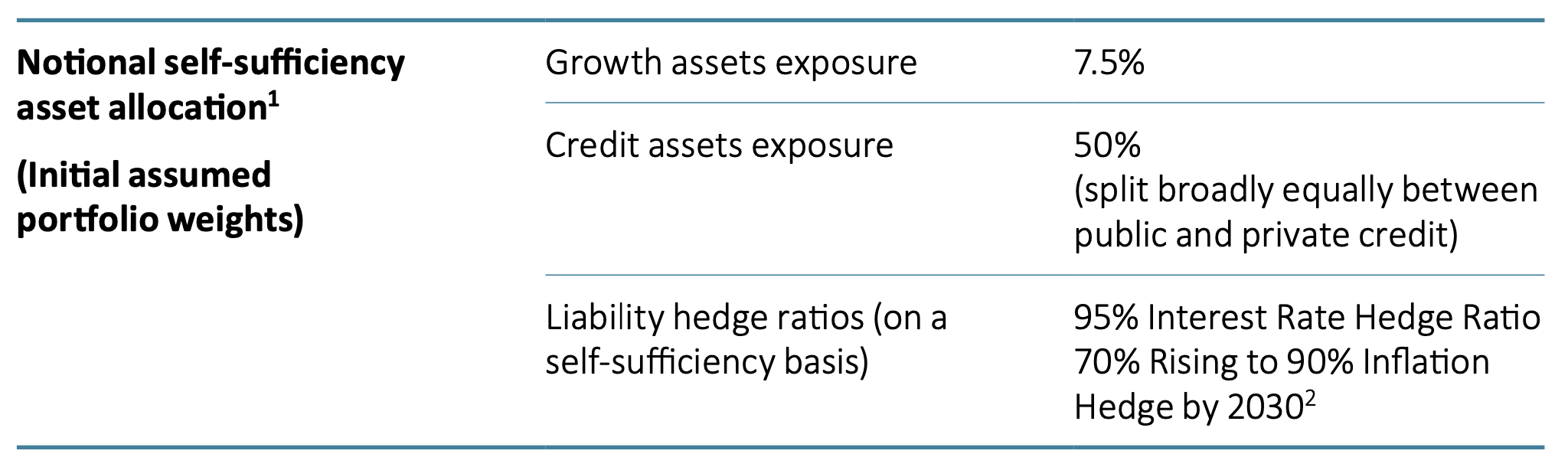}

    \label{fig:quoteSfStab}

\end{figure}
\tiny\textbf{Notes}
  1. Please note that these percentage allocations, do not add up to 100\%, because we show liability matching assets separately, in terms of their hedge ratios.
   2. The inflation hedge ratio increases to the long term target of 90\% to coincide with RPI reform

\small

\textbf{We are content that this investment strategy will support a self-sufficiency discount rate at 31 March 2023 of gilts + 0.5\%, noting the following:}

While this investment strategy and discount rate comfortably passes the capital exhaustion test with a 95\% confidence level, it does not quite pass the funding test at the same level.

However, we note that the modelling for the funding test is:

\begin{itemize}
    \item Highly sensitive to the input assumptions.
    \item Particularly binding in the early years of the simulation.
\end{itemize}

For instance, slight (and reasonable) adjustments to any of the following assumptions materially impact the results of the funding test:

\begin{itemize}
    \item The level of interest rate hedging (for example, reducing it from 95\% to 90\%).
    \item Changing the 90\% funding threshold (for example, to 87.5\% or 85\%).
    \item Changing the time taken to reach 90\% inflation hedging (for example, from 6 to 10 years).

\end{itemize}

As a result, the trustee believes it is inappropriate to lower the self-sufficiency discount rate further to precisely meet the 95\% confidence level for both tests.

This leads to a self-sufficiency liability value of \textbf{\pounds 78.2bn} (as advised by the Scheme Actuary).
\end{quote}
\normalsize
 ------------------------------------------------------------------------------------

\textbf{2020 USS SfS definition}: The 2020 SfS definition did include reference to the funding ratio condition and is quoted on the first page of Section \ref{sec:self sufficiency}.
\\
------------------------------------------------------------------------------------

\textbf{2018 USS SfS definition}: The 2018 SfS definition did not include reference to the funding ratio condition. 

\small\begin{quote}
The Trustee considers “self-sufficiency” as the amount of assets that would be required to fund the scheme using a low-risk investment portfolio – one that has less than a 5\% chance of ever requiring a further contribution from employers. This is not a target for the Trustee, but it is an important metric that provides a view on the level of risk being taken. [p24, USS Actuarial Valuation 2018 \cite{USS_vals_mon}]
\end{quote}
\normalsize

\textbf{2017 USS SfS definition}: The 2017 USS SfS definition did not include reference to the funding ratio condition. 
\small\begin{quote}
“Self-sufficiency” for the purposes of the valuation is intended to be a measure of the value of assets required by the trustee to meet all accrued pension benefits with only a “low” probability of requiring further contributions from employers. [sec 4.4.4 on pp. 24-26, TP Consultation 2017 \cite{USS_vals_2014-2017}]
\end{quote}
\normalsize
------------------------------------------------------------------------------------

\textbf{2014 USS SfS definition}: The 2014 USS SfS definition did not include reference to the funding ratio condition. 

\small\begin{quote}
Here the liabilities of the scheme are calculated using a discount rate consistent with a low investment risk approach, where a low level of reliance is placed on the participating employers to provide further financial support to the scheme. A low investment risk approach is one that could, in appropriate scenarios, be adopted by a trustee to reduce the longer term reliance on the participating employer(s) and to reduce the likelihood of the employer not being available to meet funding shortfalls. [p.38, Consultation 2014 \cite{USS_vals_2014-2017}]
\end{quote}
\normalsize
------------------------------------------------------------------------------------

\textbf{2018 USS Response Paper}: The funding ratio condition is discussed as being consistent within the parameters of Test 1 that was adopted for both the 2014 and 2017 valuations \cite{FATest12017, Marsh_2018}.

\small\begin{quote}
... outlined below is a slightly different approach [including the funding ratio condition] to testing the contention which is consistent with the parameters of Test 1 and hopefully will allow a common understanding to be reached. [p2 USS 2018 Response \cite{Otsuka_response_2018}]
\end{quote}
\normalsize
------------------------------------------------------------------------------------

\textbf{2019 Joint Expert Panel (JEP) SfS definition}: There is no mention of a funding ratio condition in either the first or second JEP reports. The JEP self-sufficiency definition only refers to ability to pay benefits. 

\small\begin{quote}
For the 2017 and 2018 valuations, the self-sufficiency liability value is the amount of money which is enough to pay for the liabilities in 95\% of modelled future scenarios assuming that a specified investment strategy is followed. The self-sufficiency basis is the set of financial and demographic assumptions used to value the self-sufficiency liability. \cite{JEP2}
\end{quote}
\normalsize

\textbf{2024 University College London (UCL) SfS definition}: Several Universities maintain USS glossaries but omit self-sufficiency. UCL seems to be the only university that includes self-sufficiency. The UCL definition only refers to ability to pay benefits and makes no reference to a funding ratio condition. It is identical to the UUK glossary definition below. 

\textbf{2024 UUK USS Employers Glossary SfS definition}: UUK maintains a glossary on their website that includes a self-sufficiency definition. According to \href{https://web.archive.org/web/20200222195019/https://www.ussemployers.org.uk/background/glossary-key-terms}{WayBackMachine} the definition has not changed since February 2020. The definition only refers to the ability to pay benefits and makes no reference to the funding ratio condition. 

\small\begin{quote}
The status a DB scheme achieves when it can rely on low-risk/low-return investments to pay all the pensions it owes, without expecting to need further contributions.\cite{UUK_glossary}.
\end{quote}
\normalsize
------------------------------------------------------------------------------------

\textbf{2018 The Pensions Regulator Guidance}: Although TPR no longer seems to refer to self-sufficiency (rather it refers to low-dependency) the 2018 TPR guidance on self-sufficiency did not seem to reference a funding ratio condition\footnote{The Pensions Act 2024, 221A(3)(a), does refer to a `funding level' at a `relevant date' in the context of benefits provided over the long term \cite{UKParl_PensionsAct_2005_S2222}.}.

\small\begin{quote}
When a pension scheme reaches a certain level of assets (the self-sufficiency level), it expects to be able to sustain itself by investing those assets on a low risk basis and pay members’ benefits as they arise without any additional support from the employer. [\href{https://realreturns.blog/2018/10/14/running-off-db-pensions-what-is-self-sufficiency/}{What is self-sufficiency?}] \end{quote}
\normalsize

\subsection{USS best estimates of returns}\label{app:best-est}

\begin{figure} [hb!]
    \centering
\includegraphics[width=0.90\textwidth]{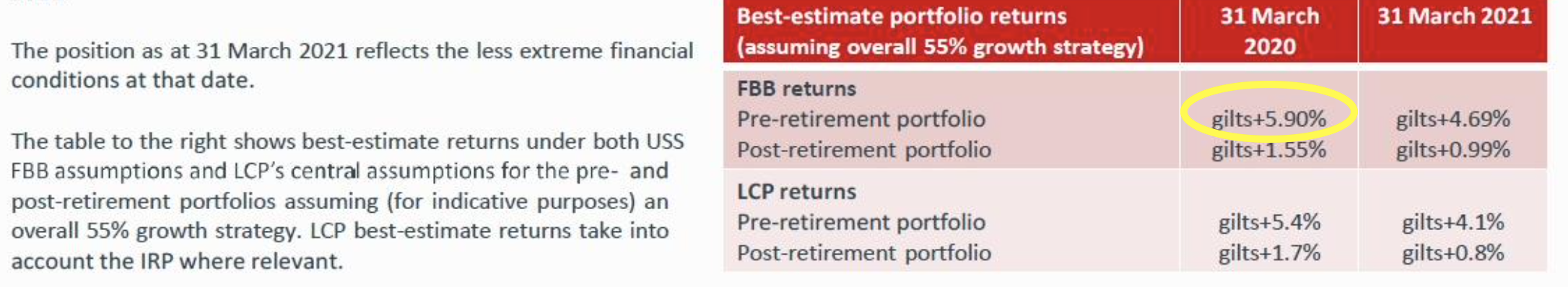}
\includegraphics[width=0.80\textwidth]{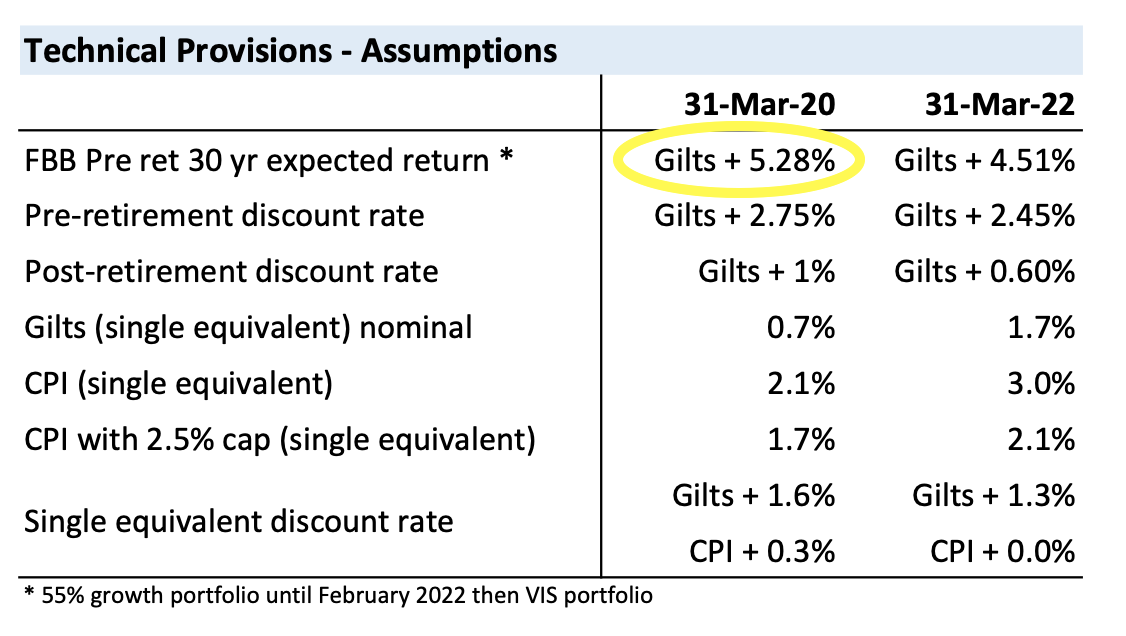}
\includegraphics[width=0.80\textwidth]{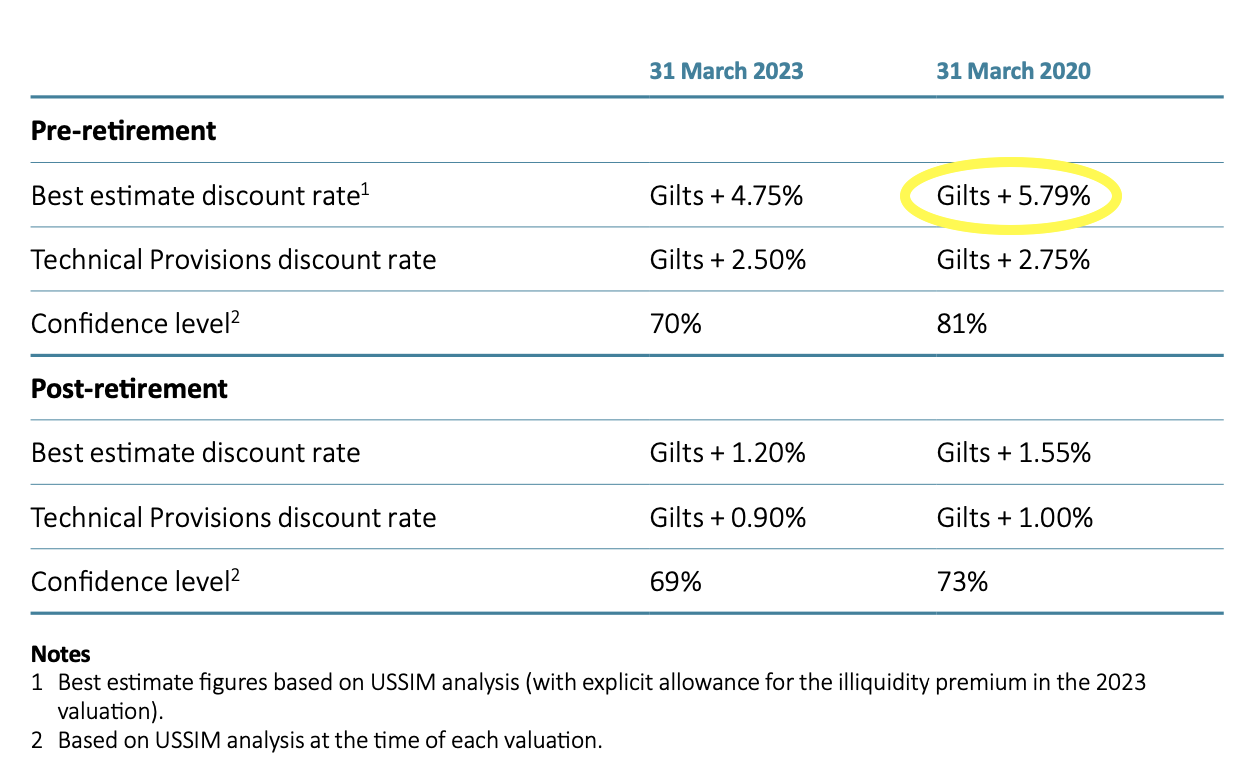}

  \caption{The March 2020 valuation best estimate of pre-retirement portfolio returns are quoted variously by USS as highlighted using yellow in the images. Top:  gilts+5.90\% in The likely outcome of a 2021 valuation, p12, July 2021 \cite{USS_Val_TP_SI_2020}. Middle: gilts+5.28\% in all Financial Management Plan post-2020 monitoring \cite{USS_vals_mon}. Bottom: gilts+5.79\% in the 2023 Technical Provisions consultation, p19 \cite{USS_Val_TP_SI_2023}.
  }
    \label{fig:BE}

\end{figure}

USS does not present their assumed or calculated best estimates of returns. Some documents do include pre- and post-retirement portfolio returns, however these can be inconsistent between documents. For example, the reported
pre-retirement portfolio best estimates for 
the March 2020 valuation
are quoted as three different values across USS documents, as reproduced in Figure \ref{fig:BE}. 
These three different values are:
\begin{enumerate}
    \item gilts+5.90\% quoted in the July 2021 `The likely outcome of a 2021 valuation',
    \item gilts+5.28\% quoted in the all post-2020 monitoring, 
     \item gilts+5.79\% quoted in the 2023 Technical Provisions consultation document.
\end{enumerate}
\newpage 

\newpage 

\subsection{Glossary}\label{glossary}

\begin{table}[H]
    \centering
\begin{tabular}[t]{|p{0.2\textwidth}|p{0.8\textwidth}|}

\hline
\textbf{Terms used} & \textbf{Description} \\
\hline
\hline
\raggedright Affordable Risk Capacity (AffRC)
& 
AffRC is defined by USS as a measure of the trustee’s and employers’ risk appetite. It aims to represent the value of the covenant. USS adopted the same approach to AffRC for 2023 and 2020. It is calculated as `Present Value of 10\% of Pensionable Payroll over 30 years +/- 5\% (with the range intended to acknowledge uncertainty).' The present value is calculated using an AffRC discount rate, and the monitoring of the AffRC discount rate shows a very high ($R^2=97\%$) correlation with the gilt yield, so the AffRC is dependent on the gilt yield as shown in Figure \ref{fig:TP Tar_Rel_Gilt} and also in the data at \cite{SussexUCUgithub}. An alternative approach is to add the covenant over time as considered in Figures \ref{fig:Benefit_payment} and \ref{fig:SfS_indABCD}. See pp4-6 of the \href{https://www.uss.co.uk/-/media/project/ussmainsite/files/about-us/valuations_yearly/2023-valuation/technical-provisions-consultation-supporting-information-19-july-2023.pdf}{2023 TP Consultation Supporting Information} for further discussion.
\\
\hline

Bonds
& 
Bonds are investments where investors, like USS, lend money for a set period of time, in exchange for regular interest payments. Bonds are often referred to as a fixed income stream. They can be issued by governments, companies and institutions including universities. See
\href{https://www.forbes.com/advisor/investing/what-is-a-bond/}{Forbes: what is a bond?} for more information.
\\
\hline
Cashflows  
& 
For USS, this is the amount of cash needed on a monthly basis to pay all accrued pension liabilities as they fall due. They are sometimes called `armadillos' as the cashflow profile in time is like an armadillo whose head is at the current year and whose tail is several decades in the future.  Cashflows can be split according the allocation of active, deferred or retired scheme members, and an example can be seen on p3 of the 
\href{ https://www.uss.co.uk/-/media/project/ussmainsite/files/about-us/our-valuation/uss-actuarial-valuation-report-2023.pdf}{USS: 2023 actuarial valuation}. USS also publish annual tables of their estimated cashflows. For example 
\href{https://www.uss.co.uk/-/media/project/ussmainsite/files/about-us/valuations_yearly/2020-valuation/2020-valuation-cashflows-for-stakeholders-update.pdf}{USS: cashflows 2020} and \href{ https://www.uss.co.uk/-/media/project/ussmainsite/files/about-us/our-valuation/uss-actuarial-valuation-report-2023.pdf}{USS: cashflows 2023}.
\\
\hline
\raggedright Contributions
& 
Contributions or contribution rates are the amount paid towards pensions as a fraction of salary. They are usually expressed as a percentage and paid on a monthly basis. They are paid partly by employers and partly by individual employees. Since 1975, for USS, the share has been around 70:30, employers to employees, see Figure 1 and linked spreadsheet of \cite{SussexUCU2023}. This 70:30 split is sometimes confused with the cost-sharing rule (see `Cost-sharing'). The total contributions include Future Service Costs and Deficit Recovery Contributions, plus additions for scheme overheads. They are detailed in the \href{https://www.uss.co.uk/about-us/valuation-and-funding/schedule-of-contributions}{USS Schedule of Contributions}. 
\\
\hline
\end{tabular}
\end{table}

\begin{table}[H]
    \centering
\begin{tabular}[t]{|p{0.2\textwidth}|p{0.8\textwidth}|}
\hline
\raggedright Cost-sharing
& 
The USS rule on cost-sharing between employers and employees \textit{in the absence of agreement} is for a 65:35 split in the \textbf{change} of contributions. This is detailed in USS Scheme Rules 64.10, and 76.4 to 76.8. See \href{https://www.uss.co.uk/about-us/scheme-rules}{USS: Scheme Rules}. It is sometimes confused with the \textit{actual agreed split} between employers and employees which has been consistently around 70:30 since 1975 (see `Contributions').
\\

\hline
Covenant
& 
According to the Pensions Regulator, the `covenant is the extent of the employer’s legal obligation and financial ability to support the scheme now and in the future.' 
\href{https://www.thepensionsregulator.gov.uk/en/document-library/scheme-management-detailed-guidance/funding-and-investment-detailed-guidance/assessing-and-monitoring-the-employer-covenant}{TPR: assessing and monitoring the employers covenant}. For USS, which is a collective employer scheme with around 340 employers, the covenant is considered to be strong and is valued at 10\% of payroll over 30 years. See also \href{https://www.uss.co.uk/about-us/valuation-and-funding/debt-monitoring-framework}{USS: debt-monitoring and covenant support}.
\\
\hline

Deficit Recovery Contributions (DRCs)
& 
DRCs are the percentage of contributions used to pay off any Technical Provisions deficit. The Pensions Regulator's Draft DB funding code states that the `deficit shown in the valuation (subject to post-valuation experience and investment outperformance adjustments, where relevant) must be recovered as soon as the employer can reasonably afford.' \href{https://www.thepensionsregulator.gov.uk/en/document-library/consultations/draft-defined-benefit-funding-code-of-practice-and-regulatory-approach-consultation/draft-db-funding-code-of-practice#reasonableaffordability}{TPR: draft funding code}. DRCs are the employers responsibility. However, if USS consider DRCs are necessary, then clearly the remaining contributions available for future pensions are reduced.  The cost-sharing rule is implemented for any \textit{change} in contributions if no agreement can be reached, and this includes deficit contributions. These arrangements effectively mean employees contribute to DRCs.
\\
\hline
\raggedright Defined Benefit (DB)  & A DB pension is `defined' or guaranteed to be a particular value every year during retirement. It can be thought of as a regular guaranteed income in retirement. The amount received is usually calculated from average salary and the length of time worked. If it is index linked, e.g. to CPI, it will retain its value in real terms. 
\\
\hline
\raggedright Discount rate (DR) & 
A discount rate is the rate of return used  to define a net present value of liabilities from the cashflows. USS estimate projected cashflows needed to pay benefits as they fall due (see `Cashflows'), then use a discount rate to estimate a net present value of pension benefits due. This gives a discount rate dependent value of liabilities as a single number in GBP.
For more details see \href{https://corporatefinanceinstitute.com/resources/valuation/discount-rate/}{Corporate Finance Institute: discount rate}. USS use a number of discount rates including: TP, FSC, SfS, AffRC (see Section \ref{sec:ActTarRel} for definition of AffRC), best estimate, break-even, pre-retirement, post-retirement and the Singe Equivalent (SE) discount rate. 
\\
\hline
\end{tabular}
\end{table}

\begin{table}[H]
    \centering
\begin{tabular}[t]{|p{0.2\textwidth}|p{0.8\textwidth}|}
\hline
\raggedright Dual Discount Rate (DDR)
& 
The DDR approach aims to combine discount rates from pre-retirement and  post-retirement portfolios in a way that reflects the membership profile. The pre-retirement portfolio is highly weighted to investments in long term growth seeking assets (equities). While the post-retirement portfolio has investments in assets with lower short-term volatility and lower long term growth (bonds). The aim is to balance the two aspects of pre- and post- retirement portfolios according to the number of members who are pre- or post-retirement. It was introduced as a possible approach in the context of USS by the Joint Expert Panel to better `allow for the long term interests of members, employers and the sector to be addressed'. See p65 of the \href{https://www.ussemployers.org.uk/news/second-report-joint-expert-panel-published}{Joint Expert Panel Report 2}.
\\
\hline
Equities
& 
Equities is the term used to refer to shares as an asset class. They have a higher return potential, but  lower short term predictability than bonds. Holding shares in a company also means influence, for example through voting rights at AGMs. See
\href{https://www.forbes.com/advisor/in/investing/what-are-equities/}{Forbes: what are equities?} for more details.
\\
\hline
\raggedright Future Service  Costs (FSCs)
& 
FSCs are the amount of money paid towards future benefits. They are stated as a percentage of salary and are the payments for pensions that are earned in the present and future. They are set at each valuation and remain constant until the next valuation. They form part of the overall contributions which may also include  Deficit Recovery Contributions, which are intended to pay off any estimated deficits. FSCs are calculated from the FSC discount rate and the cashflows. For further details also see Section \ref{sec:Pre-ret}.
\\
\hline
Gilt yield 
& 
This is the term used to refer to the yield or the return on UK government bonds. See also `Bonds' and \href{https://www.dmo.gov.uk/responsibilities/gilt-market/about-gilts/}{UK government debt office: About Gilts}.
\\
\hline
Integrated Risk Management Framework (IRMF)
& 
USS state that their `IRMF defines the Trustees approach to identifying, managing and monitoring risks that can affect the funding objectives for the DB section of the Scheme.' They also state that their Valuation Investment Strategy (VIS) must retain consistency with the IRMF. A central aspect to the IRMF is self-sufficiency. There is no single USS document detailing their IRMF. See Section \ref{sec:IRMF} and pp4-6 of the \href{https://www.uss.co.uk/-/media/project/ussmainsite/files/about-us/valuations_yearly/2023-valuation/technical-provisions-consultation-supporting-information-19-july-2023.pdf}{2023 TP Consultation Supporting Information} for further discussions. 
\\
\hline
\end{tabular}
\end{table}

\begin{table}[H]
    \centering
\begin{tabular}[t]{|p{0.2\textwidth}|p{0.8\textwidth}|}
\hline
\raggedright Joint Expert Panel (JEP) & 
The Joint Expert Panel was set up in 2018 following widespread industrial action over the proposal to close USS as a DB pension scheme. The JEP produced two reports. The first report published in September 2018 can be read here: \href{https://www.ucu.org.uk/article/9632/UCU-response-to-USS-Joint-Expert-Panel-report}{UCU response to the JEP report}. The second report published in December 2019 can be read here: \href{https://www.ucu.org.uk/article/10494/UCU-response-to-USS-Joint-Expert-Panels-second-report}{UCU response to the second JEP report}.
\\
\hline
\raggedright Pre-retirement portfolio and discount rate & 
 USS set a hypothetical pre-retirement portfolio of 90\% equities and 10\% bonds. They then estimate the annualised return and choose a prudent margin below this to calculate a pre-retirement discount rate. The pre-retirement and post-retirement discount rates combine in the Dual Discount Rate approach.  The Future Service Cost discount rate and the Technical Provisions Discount rate are then calculated from ratios of the pre- and post-retirement discount rates as described in Section \ref{sec:Pre-ret} and \ref{subsec:TP DR gilt yield}.
\\
\hline
\raggedright Post-retirement portfolio and discount rate & 
USS set a hypothetical post-retirement portfolio of 90\% bonds and 10\% equities. A post-retirement discount rate is then chosen by the same principle as for the pre-retirement discount rate. The pre-retirement and  post-retirement discount rates combine in the Dual Discount Rate approach.  The Future Service Cost discount rate and the Technical Provisions Discount rate are then calculated from ratios of the pre- and post-retirement discount rates as described in Section \ref{sec:Pre-ret} and \ref{subsec:TP DR gilt yield}.
\\
\hline
\raggedright Reliance on covenant & 
See `Covenant' and Section \ref{sec:ActTarRel}.\\
\hline
Run-off
& 
Run-off can take lots of meanings including as a reference to \href{https://www.investopedia.com/terms/r/runoff.asp}{Ticker Tape}! But in this case the meaning is that USS would be modelled as closed to new members and to new accrual. So it receives no new contributions but continues to pay those benefits accrued as they fall due. So the scheme runs down or is in run-off. Such analysis of a scheme in run-off could be attempting to answer the question: are there sufficient funds in the scheme today to pay all benefits promised so far, as they fall due? See Section \ref{sec:self sufficiency}.
\\
\hline
\raggedright  Self-sufficiency (SfS)  
& This is a term used in a specific way by USS. The 2020 valuation key terms stated that SfS was the `assets and low-risk investment strategy that provide a 95\% chance of paying all accrued benefits without the need for additional contributions, while maintaining a high funding ratio.' See Section \ref{sec:self sufficiency} and Appendix \ref{app: self-sufficiency definitions} for discussions and evolving USS definitions. USS claim their SfS definition is needed to address the low-dependency requirement of the Pensions Regulator.
\href{https://www.thepensionsregulator.gov.uk/en/document-library/consultations/draft-defined-benefit-funding-code-of-practice-and-regulatory-approach-consultation/draft-db-funding-code-of-practice#chapter4}{TPR: Draft DB funding code, Ch. 4 Low dependency}. \textit{...continued on next page.}
\\
\hline
\end{tabular}
\end{table}

\begin{table}[H]
    \centering
\begin{tabular}[t]{|p{0.2\textwidth}|p{0.8\textwidth}|}
\hline
\raggedright  Self-sufficiency (SfS)  
& \textit{...continued from previous page.} Associated terms are SfS discount rate, SfS liabilities, SfS deficit and SfS portfolio. The SfS liabilities are calculated from cashflows using a SfS discount rate. The SfS deficit is calculated from current assets minus SfS liabilities. The SfS portfolio, used in the SfS definition, appears to be the same as the post-retirement portfolio. Both have been quoted as being around 10\% growth assets and 90\% bonds, although for 2023 the SfS portfolio was 7.5\% growth assess, see Appendix \ref{app: self-sufficiency definitions}.  
\\
\hline

\raggedright Technical Provisions (TP) 
& 
TPs or TP liabilities are the net present value of the cashflows for all accrued benefits. They are calculated using a TP discount rate, which must be chosen prudently. See Section \ref{subsec:benefit payment}. The Pensions Regulator states that a `...defined benefit (DB) scheme is subject to the statutory funding objective, which means it needs to have appropriate assets to cover its accrued liabilities (known as ‘technical provisions’)'. \href{https://www.thepensionsregulator.gov.uk/en/trustees/investment-and-db-scheme-funding/valuing-your-scheme}{TPR: Statutory Funding Objective}. If the TP liabilities are less than the assets then the scheme records a TP deficit (usually simply called the deficit) and Deficit Recovery Contributions are required. The TP liabilities, and hence deficit, are usually very sensitive to the choice of TP discount rate. 
\\
\hline
\raggedright Transition Risk  ($T_{Risk}$) 
& 
USS quote Transition Risk in GBP, and its value ranges between \pounds6-8bn. It can be thought of as a USS estimated cost associated with moving to self-sufficiency. It is defined by USS as a `measure of the market risk of moving from the current investment strategy to a self-sufficiency strategy, together with the risk of life expectancies increasing (improving) faster than assumed.  USS, p4-5 \cite{USS_Val_TP_SI_2023}. See also USS Transition Risk \cite{USS_Transition_Risk_2022}. The Transition Risk is borne by employers, so effectively reduces the value of the covenant available to support the scheme. 
\\
\hline
\raggedright Universities and Colleges Employers Association (UCEA) 
& 
UCEA membership is open to UK Higher Education providers and associated organisations. Its stated aims are to be the leading voice on employment and reward matters in the UK Higher Education sector, and to support members to be employers of choice through collaboration, advocacy and expert advice. UCEA will take over responsibility from UUK as the USS employer representative in 2024.
The announcement of the transfer can be read here:  \href{https://www.ussemployers.org.uk/news/update-transfer-uss-employer-representative-responsibilities-universities-uk-ucea}{Transfer of USS Employers' representative.}
\\
\hline

\end{tabular}
\end{table}

\begin{table}[H]
    \centering
\begin{tabular}[t]{|p{0.2\textwidth}|p{0.8\textwidth}|}
\hline
\raggedright University and College Union  (UCU) 
& 
The University and College Union is the formal representative of USS scheme members. It is a Trade Union representing over 120,000 academics, lecturers, trainers, instructors, researchers, managers, administrators, computer staff, librarians, technicians, professional staff and postgraduates in universities, colleges, prisons, adult education and training organisations across the UK. The UCU website is here: \href{https://www.ucu.org.uk}{www.ucu.org.uk}.
\\
\hline
\raggedright Universities UK (UUK) 
& 
Universities UK is the formal representative of USS scheme employers. Known until 2000 as the Committee of Vice-Chancellors and Principals of the Universities of the United Kingdom (CVCP) its membership consists of over 140 vice-chancellors or principals of UK universities. The UUK website is here: \href{https://www.universitiesuk.ac.uk}{www.universitiesuk.ac.uk}. Employer representation will transferred to UCEA during 2024. 
\\
\hline
\raggedright USS Employers 
& 
USS Employers is a website (\href{https://www.ussemployers.org.uk}{www.ussemployers.org.uk}) owned and managed by UUK, 
It is used to communicate and consult on employer issues relating to USS.
\\
\hline
Valuation 
& 
A valuation aims to assess the financial health of the scheme and it is a statutory requirement to complete a full actuarial valuation every three years. According to the Pensions Regulator a scheme `...should commission a full actuarial valuation at least every 3 years. If you obtain an interim actuarial report for each intervening year, you won’t need to commission the full valuation more frequently.
The actuarial valuation must incorporate the actuary’s certification of the technical provisions calculation and the schedule of contributions.
The valuation must include the actuary’s estimate of the scheme’s solvency.
You must choose a method for calculating the scheme’s technical provisions, ie the value of benefits accrued to a particular date. You must take advice from the actuary on the differences between the methods and their impact on the scheme.' \href{https://www.thepensionsregulator.gov.uk/en/trustees/investment-and-db-scheme-funding/valuing-your-scheme}{TPR: Valuing your scheme}.
\\
\hline
\raggedright  Valuation Investment Strategy (VIS) previously the Reference Portfolio
& 
According to USS their VIS is a theoretical, but implementable, asset allocation strategy for the \pounds 70-80bn DB scheme. It serves as a guide to the implemented strategy but maintains consistency with the USS Integrated Risk Management Framework (IRFM). See 
\href{https://www.uss.co.uk/how-we-invest/our-principles-and-approach}{USS investment principles webpage} and the \href{https://www.ussemployers.org.uk/news/2020-uss-valuation-statement-investment-principles-consultation-employers}{2020 consultation with employers on the VIS}.
\\
\hline
\end{tabular}

\caption{Glossary of key terms. See also USS  \cite{USS_glossary} and UUK glossaries \cite{UUK_glossary}.  }
    \label{tab:glossary}
\end{table}

\end{document}